\documentclass[draft]{article}

\usepackage{latexsym,e-journal}

\usepackage{amssymb,amsfonts}       

\usepackage{cite}            

\begin{document}

\newcommand{\rum}{\rule{0.5pt}{0pt}}
\newcommand{\rub}{\rule{1pt}{0pt}}
\newcommand{\rim}{\rule{0.3pt}{0pt}}
\newcommand{\numtimes}{\mbox{\raisebox{1.5pt}{${\scriptscriptstyle \times}$}}}

\renewcommand{\refname}{References}

\twocolumn[%
\begin{center}
{\Large\bf Quantum Field Mechanics: Complex-Dynamical Completion of
Fundamental Physics and Its Experimental Implications
\rule{0pt}{13pt}}\par
\bigskip
Andrei P. Kirilyuk \\
{\small\it  Institute of Metal Physics,
36 Vernadsky Avenue, 03142 Kiev-142, Ukraine\rule{0pt}{13pt}}\\
\raisebox{-1pt}{\footnotesize E-mail:
kiril@metfiz.freenet.kiev.ua}\par
\bigskip\smallskip
{\small\parbox{11cm}{%
This report provides a brief review of recently developed extended
framework for fundamental physics, designated as Quantum Field
Mechanics and including causally complete and intrinsically
unified theory of explicitly emerging elementary particles, their
inherent properties, quantum and relativistic behaviour,
interactions and their results. Essential progress with respect to
usual theory is attained due to the unreduced, nonperturbative
analysis of arbitrary interaction process revealing a
qualitatively new phenomenon of dynamic multivaluedness of
interaction results and their dynamically entangled structure
giving rise to universally defined dynamic complexity and absent
in usual, always perturbative and dynamically single-valued models
with the zero value of unreduced complexity (including substitutes
used for ``complexity'', ``chaoticity'', ``self-organisation'',
etc.). It is shown how the observed world structure, starting from
elementary particles and their interactions, dynamically emerges
from the unreduced interaction between two initially homogeneous,
physically real protofields. In that way one avoids arbitrary
imposition of abstract entities and unprovable postulates or
``principles'', let alone ``mysteries'', of conventional theory,
which now obtain unified, consistent and realistic explanation in
terms of unreduced dynamic complexity. We complete the description
of fundamental world structure emergence and properties by an
outline of experimental implications and resulting causally
substantiated change of research
strategy.\rule[0pt]{0pt}{0pt}}}\bigskip
\end{center}]{%

\section{Introduction}\label{Sec:Intro}
Despite the recognised essential incompleteness and ruptures of
canonical fundamental science (see e. g.
\cite{Feynman,Dirac,Penrose:Shadows,Penrose:Intro,Holland,Hartle,
tHooft:QuantGrav,tHooft:PathInt,tHooft:DetQM,Tegmark,TegmarkWheeler,
Kent,Khren:Proc,Paty,Laloe}), its scholar, ``mainstream''
development continues in the direction of purely technical
variations of existing abstract models of reality that never solve
the existing fundamental problems and confirm thus the famous
``end-of-science'' thesis \cite{Horgan}. In that way modern
``exact'' science persists in the tradition and impasses of
so-called ``new physics'' that has appeared a hundred years ago as
an ``operational'', purely \emph{technical} arrangement of
\emph{qualitatively} new experimental facts, but since then has
failed to increase or even reproduce the internal consistency and
realism of previous, ``classical'' form of science, in spite of
hard enough efforts in that direction. After a spectacular
``revolutionary'' period at the beginning of the twentieth
century, the ``new'' fundamental physics has itself been
transformed into the next, quite conventional standard which,
however, not only accepts, but actually welcomes a growing variety
of \emph{inexplicable}, but postulated ``mysteries'',
``paradoxes'' and accumulating ``unsolvable'' problems, contrary
to the previous kind of ``well-established'' science and
suspiciously similar to quite different, openly esoteric
knowledge, such as mystical cabbala or ``experimentally
confirmed'' astrology.

That painful rupture between visible practical efficiency and
conceptual incompleteness of official science paradigm gives rise
to variously oriented attempts to find a ``definite'' problem
solution within a less contradictory system of knowledge (see e.
g. \cite{Krasnoh,PhotonDvoeg}). Such kind of work can be justified
also by a quickly growing number of \emph{practical difficulties}
in conventional science applications to more complicated systems
or deeper levels of reality, which show that the famous
``unreasonable effectiveness'' of the standard paradigm
\cite{Wigner} is quickly transformed into unjustified extension of
the first, spectacular successes of new physics application to the
\emph{simplest}, elementary systems
\cite{Kir:USciCom:ArXiv,Kir:75MatWave,Kir:100Quanta,Kir:75Wavefunc,Kir:QuMach}.
One could cite well-known cases of high-temperature
superconductivity and other many-body quantum problems with
irreducibly strong interactions and ``complex'' behaviour, among
so many other ``difficult'' problems (e.g. in cosmology) growing
in diversity and definitely escaping the ``unreasonable
effectiveness'' of standard, ``mathematically guessed'' theory.
That \emph{actually observed} tendency shows that if the very
essence of scientific method is to be preserved, including
unreduced consistency and \emph{truly} ``exact'' (objective)
nature of results, then it can be attained only within a
\emph{qualitatively new} concept, or ``paradigm'', which should
contain only the \emph{proven minimum} of accepted ``natural''
assumptions and \emph{rigorously derive} (rather than guess and
impose) its main results, always remaining in close, inseparable
\emph{entanglement with reality} considered as a \emph{system} of
\emph{all} its observed, \emph{interdependent} properties.

In this paper we propose a review of such qualitatively new,
intrinsically consistent concept and its application to simplest
physical systems, starting from elementary particles and their
interaction products. The general concept, embracing also the
\emph{causally complete}, i. e. physically and mathematically
consistent, understanding of structure and behaviour of
higher-level systems, is centred on the new, reality-based and
\emph{universally applicable} quantity of \emph{dynamic
complexity}, rigorously derived by the \emph{unreduced},
universally nonperturbative analysis of arbitrary
\emph{interaction process}
\cite{Kir:QuMach,Kir:USciCom,Kir:USymCom,Kir:SelfOrg,Kir:Fractal:1,Kir:Fractal:2,
Kir:Nano,Kir:Conscious,Kir:CommNet,Kir:SustTrans}. Its particular
realisation at the \emph{lowest complexity levels}, known as
``quantum'' behaviour, is designated as \emph{quantum field
mechanics}, including explicit, causally complete derivation of
elementary entities (space, time, particles and fields) and their
observed properties, such as space discreteness, time irreversible
flow, particle mass, charge, spin, interaction forces, quantum
measurement, quantum chaos, dynamically emerging classicality (in
a closed elementary system), and all specific features of
``quantum weirdness'' (duality, indeterminacy, uncertainty,
entanglement, complementarity), now demystified and
\emph{naturally unified} with the \emph{dynamically emerging}
``relativistic'' effects
\cite{Kir:USciCom:ArXiv,Kir:75MatWave,Kir:100Quanta,Kir:75Wavefunc,
Kir:QuMach,Kir:USciCom,Kir:DoubleSol:1,Kir:DoubleSol:2,Kir:ComDynGrav,Kir:Cosmo,
Kir:QuChaos,Kir:QuMeasure,Kir:QFM-95}.

In that way we show that all standard ``quantum mysteries'' and
``relativistic effects'' (actually postulated in usual theory)
naturally emerge as \emph{unavoidable}, totally consistent and
realistic, but nontrivial, properties of \emph{dynamically
multivalued}, or \emph{complex}, behaviour obtained, in its turn,
just by the \emph{truly exact}, unreduced analysis of underlying
interaction processes and therefore appearing also in \emph{any}
system behaviour, within the causally derived and
\emph{intrinsically unified diversity} of forms. Note that main
results of quantum field mechanics can be considered as a direct
complex-dynamical extension of the ``double solution'' concept of
Louis de Broglie
\cite{deBroglie:These,deBroglie:1927,deBroglie:1952,deBroglie:1956,
deBroglie:1959,deBroglie:1997,deBroglie:1964} almost totally
excluded from the scholar science framework, but now reappearing
as implicit, but surprisingly advanced version of unreduced
complexity at the lowest levels of world dynamics.\footnote{The
mentioned unreduced version of the double solution should be
clearly distinguished from its reduced, schematic version first
proposed by Louis de Broglie himself under the name of
``pilot-wave interpretation'', then reintroduced by Bohm
\cite{Bohm}, further developed by his followers
\cite{Holland,Bohmian:1,Bohmian:2,Bohmian:3,
Bohmian:4,Bohmian:NoChaos}, and now often presented as the unique
version of ``causal de Broglie-Bohm approach'' (see
refs.~\cite{Kir:75MatWave,Kir:DoubleSol:1,Kir:DoubleSol:2} for
more details).}

The attained causal completeness of the unreduced theory
culminates in explicit, dynamically derived \emph{unification} of
\emph{causally extended} versions of all known (correct) laws and
principles of fundamental science within a single, unified law of
\emph{conservation, or symmetry, of complexity}
\cite{Kir:QuMach,Kir:USciCom,Kir:USymCom,Kir:Conscious}, which
remains always \emph{exact} (i. e. is never ``broken'') and
\emph{intrinsically creative} (describes \emph{explicit emergence}
of new entities), properly reflects the unified harmony of natural
phenomena, and definitely eliminates the intrinsic separation,
incompleteness and skewness of conventional science laws.
Moreover, we rigorously show that the whole picture of usual
theory, including both the postulated fundamental ``mysteries''
and recently advanced unitary versions of the ``science of
complexity'', corresponds to a heavily limited, \emph{effectively
one-dimensional (or even zero-dimensional) projection} of
unreduced, \emph{dynamically multivalued (complex)} world
dynamics, which logically closes the theory by explaining
consistently both \emph{relative} successes of canonical theory
applications to the \emph{simplest} systems (i. e. the
``unreasonable effectiveness'') and its persisting failure to
complete the picture and understand more complicated system
behaviour. We obtain thus a direct, dynamically multivalued
\emph{extension} of usual, \emph{dynamically single-valued}, or
\emph{unitary}, theory, which is the expected natural way of
development, correlating with the attained causally complete
solution of accumulated fundamental and practical problems now
shown to be indeed fundamentally unsolvable \emph{within the
artificially limited framework} of conventional, unitary science
doctrine.

The basic framework and conclusions of the extended theory are
considered in section \ref{Sec:UniDynEmerge}, while various
experimental manifestations of the obtained results are summarised
in section \ref{Sec:ExpConf}, with the ensuing suggestion of
necessary, practically important changes in particular
applications and general research strategy.

\section{Unified dynamic origin of elementary\\ particles,
their intrinsic properties,\\ quantum and relativistic
behaviour}\label{Sec:UniDynEmerge}
Any real system structure and existence is determined by
interaction between its components. The simplest, least structured
initial configuration of a system is represented by two
effectively homogeneous and infinite (``omnipresent'') fields
attracted to each other. Guided by a strict version of Occam's
principle of parsimony, we should accept such interaction
configuration as the basis of our world construction, provided
that we can demonstrate \emph{explicit emergence} of \emph{all}
its main structures and verified laws (patterns of behaviour) in
that interaction process. Any ``postulated'' addition can be
accepted only within a proven, necessary \emph{minimum} for
observed property appearance, so that one avoids the ambiguity of
arbitrary, subjectively ``convenient'' postulates of conventional,
``positivistic'' science that only fix the facts, but explain
nothing at all. Well-specified realism and consistency of the
assumed system configuration are additionally supported by the
observed universality of \emph{two and only two} distributed
interaction forces, electromagnetic (e/m) and gravitational ones,
which means that one of the protofields will eventually be
responsible for e/m and another for gravitational interactions
(see the end of this section). It means also that our interacting
protofields are \emph{physically real} fields, even though one can
directly observe within this world only their interaction-driven
\emph{perturbations} forming the world structure. Being physically
real entities, the protofields have their own internal structure
(though usually inaccessible to \emph{direct} observation) and
related mechanical properties, including finite compressibility.

During interaction the e/m protofield degrees of freedom, $q$, are
mixed, or ``entangled'', with those of gravitational protofield,
$\xi$, into a generally ``nonseparable'' combination, $\Psi \left(
{\xi,q } \right)$, called \emph{state-function} and describing the
emerging world structure (in reality it measures the magnitude of
\emph{e/m} protofield perturbations, while a more inert, ``heavy''
gravitational medium is always ``hidden'' behind them and shows
itself only through indirect effects, such as interaction between
emerging entities, as it is described below). The state-function
$\Psi \left( {\xi,q } \right)$ obeys a dynamic equation describing
protofield interaction that will be called here \emph{existence
equation}. According to our rule of maximum self-consistency (or
minimum assumptions), one has no right to use any particular
``model'' or postulated dynamical ``principle'' for that equation,
and therefore it is accepted in the form that fixes only the fact
of protofield interaction as such:
\begin{equation}\label{Eq:ExistProto}
\left[ {h_{\rm g} \left( \xi  \right) + V_{{\rm eg}} \left( {\xi,q }
\right) + h_{\rm e} \left( q \right)} \right]\Psi \left( {\xi,q }
\right) = E\Psi \left( {\xi,q } \right),
\end{equation}
where $h_{\rm e} \left( q \right)$ and $h_{\rm g} \left( \xi
\right)$ are ``generalised Hamiltonians'' for free
(non-interacting) protofields (i. e. measurable functions
eventually expressing dynamic complexity defined below), $V_{{\rm
eg}} \left( {\xi,q } \right)$ is an arbitrary (though eventually
attractive and binding) interaction potential between the fields
of $q$ and $\xi$, and $E$ is energy (generalised Hamiltonian
eigenvalue). Being understood in a large sense,
eq.~(\ref{Eq:ExistProto}) can correspond, up to minor details, to
practically any particular ``model'' and actually reflects only
the fact of protofield interaction as such. Its ``Hamiltonian''
form is also confirmed self-consistently by further interaction
analysis
\cite{Kir:QuMach,Kir:USciCom,Kir:USymCom,Kir:Conscious,Kir:CommNet,Kir:SustTrans,
Kir:DoubleSol:1,Kir:DoubleSol:2,Kir:ComDynGrav,Kir:Cosmo}.

Let us note from the beginning the intrinsically ``cosmological''
nature of the theory formulation, where the demand of
\emph{explicit dynamic emergence} of \emph{any} entity, together
with its properties (expressed by ``laws''), ensures that our
derivation follows the main lines of real world structure
emergence and evolution. In particular, there is no inserted
``space'', ``time'', or any other observed entities and properties
in eq.~(\ref{Eq:ExistProto}): they \emph{will be obtained} from
its \emph{unreduced solution}. On the other hand, one does need to
have a primordial interaction behind the emerging world structure,
in its \emph{minimal} and correspondingly \emph{irreducible}
version, as opposed to the corrupt idea of ``emergence from
nothing'' of finally \emph{postulated} structures, underlying the
growing ``difficulties'' of official cosmology and related to the
general fallacy of ``development without change'' of unitary
science (see \cite{Kir:USciCom,Kir:Cosmo} and discussions below).

For a more convenient analysis of interaction process, we express
$\Psi \left( {\xi,q } \right)$ in terms of complete system of
eigenfunctions, $\left\{ \phi _n \left( q \right) \right\}$, of
the free e/m protofield Hamiltonian, $h_{\rm e} \left( q \right)$:
\begin{equation}\label{Eq:StFuncExpan}
\begin{array}{cc}
\Psi \left( {\xi,q } \right) = \sum\limits_n {\psi _n \left( \xi
\right)} \phi _n \left( q \right), \\[+12pt]
h_{\rm e} \left( q \right)\phi _n \left( q \right) = \varepsilon _n
\phi _n \left( q \right),
\end{array}
\end{equation}
which transforms eq.~(\ref{Eq:ExistProto}) into a system of
equations:
\[
\left[ {h_{\rm g} ( \xi  ) + V_{00} ( \xi )} \right]\psi _0 ( \xi
) + \sum\limits_n {V_{0n} } ( \xi )\psi _n ( \xi ) = \eta \psi _0
( \xi ), \hspace {0.3 cm}
\]
\begin{equation}\label{Eq:ExistSystem}
\left[ {h_{\rm g} ( \xi ) + V_{nn} ( \xi )} \right]\psi _n ( \xi )
+ \sum\limits_{n' \ne n} {V_{nn'}} ( \xi )\psi _{n'} ( \xi ) =
\hspace {0.7 cm}
\end{equation}
\[
= \eta _n \psi _n \left( \xi \right) - V_{n0} \left( \xi
\right)\psi _0 \left( \xi \right), \hspace {0.4 cm}
\]
where $\eta _n  \equiv E - \varepsilon _n$,
\[
V_{nn'} \left( \xi  \right) = \int\limits_{ \Omega _q } {dq} \phi
_n^ *  \left( q \right)V_{{\rm eg}} \left( {\xi,q } \right)\phi
_{n'} \left( q \right),
\]
and we have separated the equation with $n = 0$, while assuming
that $n \ne 0$ in other equations (here and below) and designating
$\eta \equiv \eta _0$. Note that the obtained system of equations,
eqs.~(\ref{Eq:ExistSystem}), is just another, more relevant form
of existence equation, eq.~(\ref{Eq:ExistProto}), that does not
involve any additional assumption or result of interaction
development. Such expression of system configuration in terms of
dynamical ``eigen-modes'' (or ``elements'') of a ``free''
component should always be possible, including the case of the
formally ``nonlinear'' protofields, since we suppose that the
internal dynamics of the latter is known, or ``integrable'', at
least in the scale range of interest (and does not involve itself
the observed structure emergence).

Expressing $\psi_n(\xi)$ through $\psi_0(\xi)$ from
eqs.~(\ref{Eq:ExistSystem}) with the help of standard Green
function technique \cite{Dederichs,Kir:Channel} and inserting the
result into the equation for $\psi_0(\xi)$, we reformulate the
problem in terms of \emph{effective} existence equation containing
only gravitational variables ($\xi$):
\begin{equation}\label{Eq:ExistEff}
\left[ {h_{\rm g} \left( \xi  \right) + V_{{\rm eff}} \left( {\xi
;\eta } \right)} \right]\psi _0 \left( \xi  \right) = \eta \psi _0
\left( \xi  \right),
\end{equation}
where the \emph{effective potential (EP)}, $V_{{\rm
eff}}(\xi;\eta)$, is given by
\begin{eqnarray}
V_{{\rm eff}} \left( {\xi ;\eta } \right) = V_{00} \left( \xi
\right) + \hat V\left( {\xi ;\eta } \right), \hspace {1.4 cm} \label{Eq:EP} \\[+4pt]
\hat V\left({\xi ;\eta } \right)\psi _0 \left( \xi  \right) =
\int\limits_{ \Omega _\xi  } {d\xi 'V\left( {\xi ,\xi ';\eta }
\right)} \psi _0 \left({\xi '} \right), \hspace {0.6 cm} \nonumber\\[+4pt]
\hspace {0.5 cm} V\left( {\xi ,\xi ';\eta } \right) =
\sum\limits_{n,i} {\frac{{V_{0n} \left( \xi  \right)\psi _{ni}^0
\left( \xi \right)V_{n0} \left( {\xi '} \right)\psi _{ni}^{0*}
\left( {\xi '} \right)}}{{\eta  - \eta _{ni}^0  - \varepsilon
_{n0} }}}\ , \label{Eq:EP:Kern}
\end{eqnarray}
$\varepsilon _{n0}  \equiv \varepsilon _n  - \varepsilon _0$, and
$\{\psi_{ni}^0(\xi)\}$, $\{\eta_{ni}^0\}$ are the complete sets of
eigenfunctions and eigenvalues for an auxiliary, truncated system of
equations (where $n,n' \ne 0$):
\begin{equation}\label{Eq:AuxSyst}
\left[ {h_{\rm g} ( \xi  ) + V_{nn} ( \xi )} \right]\psi _n ( \xi
) + \sum\limits_{n' \ne n} {V_{nn'} ( \xi )} \psi _{n'} ( \xi ) =
\eta _n \psi _n ( \xi ).
\end{equation}
The general solution of existence equation,
eq.~(\ref{Eq:ExistProto}), is then obtained as
\cite{Kir:QuMach,Kir:USciCom,Kir:DoubleSol:1,Kir:ComDynGrav}:
\begin{eqnarray}
{ \Psi} \left( {\xi,q } \right) = \sum\limits_i {c_i } \left[ {\phi
_0 \left( q \right) + \sum\limits_n {\phi _n } \left( q \right)\hat
g_{ni} \left( \xi  \right)} \right]\psi _{0i}
\left(\xi  \right), \label{Eq:StateFunc} \\[+4pt]
\psi _{ni} \left( \xi  \right) = \hat g_{ni} \left( \xi
\right)\psi _{0i} \left( \xi  \right) \equiv \int\limits_{ \Omega
_\xi  } {d\xi 'g_{ni} \left( {\xi ,\xi '} \right)\psi _{0i}
\left({\xi '} \right)},\nonumber\\[+4pt]
g_{ni} \left( {\xi , \xi '} \right) = V_{n0} \left( {\xi '}
\right)\sum\limits_{i'} {\frac{{\psi _{ni'}^0 \left( \xi
\right)\psi _{ni'}^{0*} \left( {\xi '} \right)}}{{\eta _i  - \eta
_{ni'}^0  - \varepsilon _{n0} }}}\ ,  \label{Eq:StateFunc:Kern}
\hspace {0.5 cm}
\end{eqnarray}
where $\{\psi_{0i}(\xi)\}$ are the eigenfunctions and $\{\eta_i\}$
eigenvalues of the effective existence equation,
eq.~(\ref{Eq:ExistEff}), while coefficients $c_i$ should be
determined from the state-function matching conditions along the
boundary where interaction vanishes. The observed system density,
$\rho(\xi,q)$, is given by the squared modulus of protofield
amplitude, $\rho(\xi,q) = \left|{ \Psi}(\xi,q)\right| ^2$.

The obtained problem expression and formal ``solution'',
eqs.~(\ref{Eq:ExistEff})--(\ref{Eq:StateFunc:Kern}), are known
under the name of optical, or effective, potential method
\cite{Dederichs}, which is usually applied, however, in a
perturbatively reduced version, where the ``nonintegrable''
self-consistent links in EP and solution expressions to unknown
eigenvalues and eigenfunctions are eliminated by one or another
approximation transforming the effective problem solution into a
``closed'', integrable form. We show that such reduction actually
\emph{kills} the \emph{qualitatively} important features of
unreduced solution that provide a \emph{universal source} of
\emph{dynamic randomness}, \emph{entanglement} of interaction
components and related \emph{squeeze (or ``reduction'') and
extension cycles}, leading to universally defined dynamic
complexity of unreduced interaction process and the unified,
causally complete understanding of elementary particles and all
their ``quantum'' and ``relativistic'' properties.

The most important property of unreduced solution revealed by its
EP expression,
eqs.~(\ref{Eq:ExistEff})--(\ref{Eq:StateFunc:Kern}), is called
\emph{dynamic multivaluedness}, or \emph{redundance}, and provides
the \emph{universal, dynamic origin of randomness} appearing as
irreducible ``quantum indeterminacy'' at the level of elementary
particle dynamics. It is due to the unreduced EP dependence on the
eigenvalues to be found, $\eta$, expressing the \emph{essential},
\emph{dynamically emerging} problem \emph{nonlinearity}, which is
not \emph{explicitly} present in its initial formulation,
eqs.~(\ref{Eq:ExistProto}),\,(\ref{Eq:ExistSystem}), that can well
contain \emph{formally linear} equations. Indeed, if $N_{\rm{e}}$
and $N_{\rm{g}}$ are the numbers of participating eigen-modes (or
``elements'') of free e/m and gravitational protofields
respectively (normally $N_{\rm{e}}  = N_{\rm{g}}$ and $N_{\rm{e}}
,N_{\rm{g}}  \gg 1$), then the total number of eigenvalues for
eqs.~(\ref{Eq:ExistEff})--(\ref{Eq:EP:Kern}), determined by the
maximum power of characteristic equation, is easily estimated as
\begin{equation}\label{Eq:FullSolutionNumber}
N_{{\rm{eg}}}  = N_{\rm{g}} N_{{\rm{eg}}}^{\rm{1}}  = N_{\rm{g}}
\left( {N_{\rm{e}} N_{\rm{g}}  + 1} \right) = \left( {N_{\rm{g}} }
\right)^2 N_{\rm{e}}  + N_{\rm{g}} \ ,
\end{equation}
where the factor of $N_{{\rm{eg}}}^{\rm{1}}  = N_{\rm{e}}
N_{\rm{g}}  + 1$ is due to EP dependence on $\eta$,
eq.~(\ref{Eq:EP:Kern}), leading to the  $N_{\rm{g}}$-fold
\emph{redundance} of ordinary eigen-solution sets plus a separate,
reduced set of eigen-solutions that forms a specific
``intermediate'' state specified below. A detailed study,
including ``geometric'' analysis of
eqs.~(\ref{Eq:ExistEff})--(\ref{Eq:EP:Kern}), confirms
universality of dynamic multivaluedness and \emph{physical
reality} of redundant partial solutions called system
\emph{realisations} because each of them is ``locally'' complete
and describes exhaustively a real system configuration
\cite{Kir:QuMach,Kir:USciCom,Kir:DoubleSol:1,Kir:ComDynGrav,
Kir:Cosmo,Kir:QuChaos,Kir:QuMeasure,Kir:Channel}.

All realisations are real and have ``equal rights'' to appear as a
result of interaction development. But since each realisation is
complete as such, the system can take \emph{only one} realisation
at a time, i. e. realisations are \emph{mutually incompatible} and
cannot be ``superimposed'', ``coexist'', or appear
simultaneously.\footnote{This is a fundamental difference of our
results from popular unitary \emph{imitations of complexity},
where different ``attractors'', or ``unstable orbits'', or other
empirically (or numerically) guessed ``states'' \emph{coexist} in
an \emph{abstract} ``space'' of \emph{continuously developing}
trajectories, with a \emph{fixed}, postulated system
configuration.} Therefore the system is ``forced'' (by the driving
interaction alone) to take and \emph{permanently change} its
realisations in a \emph{dynamically random} order, where
\emph{randomness} itself, and thus also \emph{chaos}, obtains its
\emph{universal, intrinsic origin} and definition as the above
dynamic multivaluedness of (any) unreduced interaction process. We
can present this result in the form of genuine, \emph{really
complete} and \emph{therefore} probabilistic \emph{general
solution} of a problem expressing the observed density $\rho
\left( {\xi,q} \right)$, as a \emph{causally probabilistic sum} of
respective densities, $\rho _r \left( {\xi,q} \right) = \left|
{\Psi _r \left( {\xi,q} \right)} \right| ^2$, for individual
realisations numbered by index $r$ here and below:
\begin{equation}\label{Eq:ProbabSum}
\rho \left( {\xi,q} \right) = \sum\limits_{r = 1}^{N_\Re} {^\oplus
\,\rho _r \left( {\xi,q} \right)} \ ,
\end{equation}
where $N_\Re $ ($ = N_{\rm{g}}$ in our case) is the total
realisation number, and the sign $\oplus$ serves to designate a
special, \emph{dynamically probabilistic} meaning of the sum. The
latter implies that regular realisations \emph{appear and
disappear}, together with their \emph{causally derived} densities
$\rho _r \left( {\xi,q} \right)$ (see also below), in a
\emph{dynamically random}, ``spontaneous'', ``unpredictable'', or
\emph{chaotic} order, \emph{whatever} is the time of observation
and the number of registered events (an \emph{event} is
\emph{rigorously and universally obtained} now as totally dynamic,
interaction-driven \emph{realisation emergence and
disappearance}). The dynamically probabilistic sum of the general
solution, eq.~(\ref{Eq:ProbabSum}), includes, in its complete
form, further levels of chaotically changing realisations obtained
by application of the same EP method to the system of auxiliary
equations, eqs.~(\ref{Eq:AuxSyst}), and constituting
hierarchically organised branches of \emph{extended, dynamically
probabilistic fractal}
\cite{Kir:USciCom,Kir:Fractal:2,Kir:Conscious} that represents the
unreduced, complex-dynamical result of interaction development and
the emerging system structure.

Moreover, we obtain the \emph{a priory determined}, causal
(dynamic) probability, $\alpha _r$, of the $r$-th
\emph{elementary} realisation emergence as
\begin{equation}\label{Eq:ElemProbab}
\alpha _r  = \frac{1}{{N_\Re}} \ \ \left( {r = 1,...,N_\Re }
\right) \,, \ \ \ \ \sum\limits_{r = 1}^{N_\Re  } {\alpha _r }  =
1 \ ,
\end{equation}
Since in many cases elementary realisations are not individually
resolved in actual measurements and, being inhomogeneously
distributed, appear in dense groups, or dynamic ``tendencies'',
containing various their numbers, the probabilities of appearance
of such \emph{compound} realisations are causally/dynamically
determined by the numbers of constituent elementary realisations:
\begin{equation}\label{Eq:RealProbab}
\alpha _r \left( {N_r } \right) = \frac{{N_r }}{{N_\Re }}\ \left(
{N_r  = 1, ... ,N_\Re  ;\ \sum\limits_r {N_r }  = N_\Re } \right),
\end{equation}
$N_r$ being the number of elementary realisations in their $r$-th
group forming an actually observed, compound realisation. When the
number of events (observation time) is large enough, the
probabilistic sum of eq.~(\ref{Eq:ProbabSum}) approaches usual
\emph{expectation (average) value},
\begin{equation}\label{Eq:ExpValue}
\rho _{{\rm{ex}}} \left( {q,\xi } \right) = \sum\limits_{r =
1}^{N_\Re  } {\alpha _r \rho _r \left( {q,\xi } \right)} \ .
\end{equation}
However, the dynamically probabilistic sum of
eq.~(\ref{Eq:ProbabSum}) and associated causal probability values,
eqs.~(\ref{Eq:ElemProbab})--(\ref{Eq:RealProbab}), preserve their
meaning also for \emph{single, isolated} events emerging in
\emph{real time} (and thus \emph{together} with causally specified
time itself, see below).

The second fundamental property of unreduced interaction result,
the autonomous \emph{dynamic entanglement} of interacting system
components (two protofields in our case) within each realisation
is specified in the obtained solution, eq.~(\ref{Eq:StateFunc}),
by the dynamically weighted sum of products of functions of
interacting degrees of freedom, $q$ and $\xi$ (this structure is
reproduced at further developing levels of the dynamically
probabilistic fractal
\cite{Kir:USciCom,Kir:Fractal:2,Kir:Conscious}). The general
interaction result, eq.~(\ref{Eq:ProbabSum}), emerges thus as
\emph{dynamically multivalued entanglement} of system components,
where entanglement as such determines the physical, tangible
\emph{quality}, or ``texture'', of really new, emerging entities
represented by system realisations, and similar to the above
property of multivaluedness-chaoticity, that process of \emph{new
quality} formation by (fractal) dynamic entanglement is
\emph{fundamentally absent} in the unitary projection of
conventional theory.

The process of realisation formation and change can be better
understood if we specify the expressions for the state-function
(density) and EP for an arbitrary, $r$-th realisation, by
substituting its eigenvalues and eigenfunctions found from the
effective existence equation, eq.~(\ref{Eq:ExistEff}), for their
general designation in
eqs.~(\ref{Eq:EP})--(\ref{Eq:EP:Kern}),\,(\ref{Eq:StateFunc})--(\ref{Eq:StateFunc:Kern}):
\begin{equation}\label{Eq:EP-Full}
V_{{\rm{eff}}} \left( {\xi ;\eta _i^r } \right)\psi _{0i}^r \left(
\xi  \right) = V_{00} \left( \xi  \right)\psi _{0i}^r \left( \xi
\right) +
\end{equation}
\[
+ \sum\limits_{n, i'} {\frac{{V_{0n} \left( \xi \right)\psi
_{ni'}^0 \left( \xi  \right)\int\limits_{\Omega _\xi } {d\xi '\psi
_{ni'}^{0*} \left( {\xi '} \right)V_{n0} \left( {\xi '}
\right)\psi _{0i}^r \left( {\xi '} \right)} }}{{\eta _i^r  - \eta
_{ni'}^0  - \varepsilon _{n0} }}}\ ,
\]

\begin{equation}\label{Eq:StFunc-Full}
\Psi _r \left( {\xi,q} \right) = \sum\limits_i {c_i^r } \left[
{\phi _0 \left( q \right)\psi _{0i}^r \left( \xi \right)} \right.
+
\end{equation}
\[
+ \sum\limits_{n, i'} {\left. {\frac{{\phi _n \left( q \right)\psi
_{ni'}^0 \left( \xi \right)\int\limits_{\Omega _\xi } {d\xi '\psi
_{ni'}^{0*} \left( {\xi '} \right)V_{n0} \left( {\xi '}
\right)\psi _{0i}^r \left( {\xi '} \right)} }}{{\eta _i^r  - \eta
_{ni'}^0  - \varepsilon _{n0} }}} \right] }\ ,
\]
\[
\rho _r \left( {\xi,q} \right) = \left| {\Psi _r \left( {\xi,q}
\right)} \right| ^2 \ .
\]
It is not difficult to see
\cite{Kir:QuMach,Kir:USciCom,Kir:Fractal:2,Kir:DoubleSol:1,
Kir:ComDynGrav,Kir:QuChaos,Kir:QuMeasure} that because of resonant
eigenvalue involvement in the denominators of unreduced EP and
state-function expressions,
eqs.~(\ref{Eq:EP-Full}),\,(\ref{Eq:StFunc-Full}), each state
function and EP realisation concentrates around certain eigenvalue
$\eta _i^r$ that can be interpreted as a centre of emerging
elementary particle and \emph{space} structure (physical space
``point''), or its dynamically emerging ``coordinate'', while the
difference of such eigenvalues for consecutively taken,
``neighbouring'' realisations, $\Delta \eta _i^r$, forms the
elementary, discrete \emph{space distance} (or
length).\footnote{Note also the ``cutting'' role of ``overlap
integrals'' in the numerators of
eqs.~(\ref{Eq:EP-Full}),\,(\ref{Eq:StFunc-Full}) and a convenient
choice of eigenfunctions of the initial expansion basis in
eq.~(\ref{Eq:StFuncExpan}), $\left\{ {\phi _n \left( q \right)}
\right\}$, in the form of highly localised, $\delta$-like
functions.} It is important that due to the self-consistent
dynamic relation between the state-function and EP structures,
eqs.~(\ref{Eq:EP-Full}),\,(\ref{Eq:StFunc-Full}), they concentrate
around the same emerging space point, mutually amplifying each
other, i. e. taking each realisation, the system autonomously
``digs up the potential well for itself'' and
\emph{simultaneously} ``falls'' into it, without any inhomogeneity
being artificially introduced into the uniform initial
configuration. This essential \emph{dynamic instability} of
unreduced protofield interaction with respect to ``spontaneous''
\emph{dynamical squeeze, or collapse, or reduction} around
\emph{randomly} chosen realisation centre is closely related to
dynamic multivaluedness and (fractal) entanglement, where the
former gives \emph{causal} (dynamic) randomness and the latter
drives the collapse and accounts for its self-amplifying,
``catastrophic'' development.

Due to finite protofield compressibility, local reduction process
around a current realisation centre will stop at the moment of
maximum possible squeeze. But the protofield interaction persists
around the point of maximum squeeze (also due to the
\emph{dynamically fractal} interaction structure), and therefore
at a moment where the \emph{dynamic} amplification force of
collapse vanishes, the same intrinsic instability that gives rise
to reduction by self-amplified entanglement will cause the
reverse, also self-amplifying \emph{disentanglement} and
\emph{extension} process, leading to a quasi-homogeneous system
state close to initial configuration of noninteracting protofields
followed by another collapse to the next, also randomly chosen
reduction centre, and so on. That is the detailed, dynamically
inevitable mechanism of realisation change in any unreduced
interaction
\cite{Kir:QuMach,Kir:USciCom,Kir:Fractal:1,Kir:Fractal:2,
Kir:Conscious,Kir:CommNet,Kir:SustTrans,Kir:DoubleSol:1,
Kir:ComDynGrav,Kir:Cosmo,Kir:QuChaos,Kir:QuMeasure}. One obtains
thus an infinite series of cycles of protofield
reduction-extension around randomly chosen reduction centres
forming the dynamical space structure. We designate this
\emph{spatially chaotic, essentially} nonlinear pulsation in the a
priori homogeneous system of interacting protofields as
\emph{quantum beat} process and argue that it constitutes the
\emph{complex-dynamical internal structure}, and the very essence,
of any \emph{massive elementary particle}, exemplified by the
electron as the simplest species.

Note the role of transient, extended state of quasi-free
components during system transition between two successive
localised realisations: it constitutes a specific,
``intermediate'' (or ``main'') realisation, which has appeared in
the unreduced solution analysis (in relation to
eq.~(\ref{Eq:FullSolutionNumber})), enters explicitly that
solution [20-25], and forms the unified dynamical ``link'' and
probabilistically structured ``distribution ground'' for all
other, regular, localised realisations. As confirmed below, that
intermediate realisation forms a \emph{physically real},
\emph{causally complete} version of the famous quantum-mechanical
\emph{wavefunction}, possessing all the necessary properties
(``inexplicable'' in unitary theory) and universally extendible to
higher complexity levels (see also
refs.~\cite{Kir:75Wavefunc,Kir:QuMach,Kir:USciCom,Kir:Fractal:2,
Kir:Conscious,Kir:CommNet,Kir:SustTrans,Kir:DoubleSol:1,Kir:DoubleSol:2,
Kir:ComDynGrav,Kir:Cosmo}).

In particular, \emph{wave-particle duality} is \emph{dynamically}
obtained in the form of permanent change between the squeezed, or
``corpuscular'', and extended, or ``undular'', states of the
quantum beat process within any (massive) elementary particle that
can also be called \emph{field-particle} (that dynamic alternation
of the two kinds of state is naturally reproduced and observed in
interaction processes of elementary field-particles, see
below).\footnote{The realistic, complex-dynamical mechanism of
quantum duality is naturally extended to the causally complete
interpretation of a more general concept of ``complementarity''
introduced by Niels Bohr and known for its particular obscurity,
even with respect to other ``quantum mysteries''. Namely,
\emph{causal complementarity} refers to the intrinsically
dualistic nature of unreduced, \emph{multivalued interaction
dynamics} unifying, contrary to its single-valued projection, such
contradictory properties as locality (``structures'') and
nonlocality (``transitions''), continuity and discreteness
(quantization into realisations), regularity (configuration of
each realisation) and randomness (realisation emergence and
sequence), entanglement (mixture) and disentanglement
(separation), quanticity and classicality (see
refs.~\cite{Kir:QuMach,Kir:USciCom,Kir:Fractal:1,Kir:Fractal:2,
Kir:Conscious,Kir:CommNet,Kir:SustTrans,Kir:DoubleSol:1,Kir:DoubleSol:2,
Kir:ComDynGrav,Kir:Cosmo,Kir:QuChaos,Kir:QuMeasure} for more
details).}

The \emph{intrinsically probabilistic nature} of the wavefunction,
remaining its most ``mysterious'' property in all unitary theory
versions (especially in combination with its quite \emph{real}
manifestations, such as diffraction), also appears as inevitable,
dynamic consequence of underlying unreduced, multivalued
interaction process, where the canonical \emph{Born's probability
rule}, used for practical calculation of regular realisation
probabilities, follows directly from the dynamical ``matching
conditions'' for the wavefunction and its dynamically squeezed
transform, eq.~(\ref{Eq:StFunc-Full})
\cite{Kir:QuMach,Kir:USciCom,
Kir:Conscious,Kir:DoubleSol:1,Kir:ComDynGrav,Kir:Cosmo,Kir:QuChaos,Kir:QuMeasure}.
The wavefunction represents natural dynamical averaging between
regular realisations performed, in agreement with the expectation
value expression of eq.~(\ref{Eq:ExpValue}), in the
``intermediate'' realisation by the unreduced system dynamics
itself. It is \emph{only that}, ``averaged'', or ``main'',
realisation that remains in usual, dynamically single-valued
(unitary) description, while its intrinsic dynamical links to
permanently appearing and disappearing ``regular'' (localised)
realisations, as well as those realisations themselves, are
arbitrarily thrown off (and then \emph{artificially} reinserted,
in the form of ``corpuscular properties'' to ensure ``agreement
with experiment'', but at the expense of postulated ``quantum
mysteries'', ``relativistic paradoxes'', and irreducible
separation from reality).

A realisation \emph{change} (reduction-extension) cycle
constitutes the causally specified and universal essence of
physical \emph{event}, where the \emph{qualitative system
transformation}, in the form of its interaction-driven, highly
nonlinear reconstruction, plays a crucial role and shows why
events, and with them \emph{physical time}, cannot be consistently
introduced in conventional, dynamically single-valued projection
of reality. Since we have \emph{explicitly obtained} the
well-specified sequence of \emph{essentially nonlinear} events of
\emph{unceasing, qualitative, and spatially chaotic change}, we
obtain also the consistent entity of \emph{real physical time}
measured by \emph{intensity} (appearing as \emph{frequency}) of
\emph{realisation change} process and possessing the
\emph{inherent, dynamically derived} property of
\emph{irreversibility} with its two aspects of (1) \emph{unceasing
flow} (realisation \emph{change}) and (2) \emph{unpredictability
(randomness)} of each next step (i. e. \emph{time flow direction}
towards \emph{growing total randomness} naturally identified with
\emph{growing dynamic entropy}).

In that way we have obtained the minimum necessary basis for the
consistent introduction of unreduced \emph{dynamic complexity},
$C$, which can now be \emph{universally} defined
\cite{Kir:QuMach,Kir:USciCom,Kir:USymCom,Kir:SelfOrg,Kir:Fractal:1,Kir:Fractal:2,
Kir:Nano,Kir:Conscious,Kir:CommNet,Kir:SustTrans,Kir:DoubleSol:1,Kir:DoubleSol:2,
Kir:ComDynGrav,Kir:Cosmo,Kir:QuChaos} as a growing function of
system realisation number, $N_\Re$, or rate of their change, equal
to zero for the (unrealistic) case of only one realisation: $C =
C_0 f\left( {N_\Re} \right)$, where ${{df} \mathord{\left/
{\vphantom {{df} {dN_\Re   > 0}}} \right.
 \kern-\nulldelimiterspace} {dN_\Re   > 0}}$ and $f\left( 1 \right) = 0$;
for example, $f\left( n \right) = \ln \left( n \right)$ (this
definition remains valid only if the unreduced, dynamically
multivalued structure of a system or interaction process is
explicitly obtained with all its internal dynamical links).

The above \emph{explicitly derived} spatial structure and temporal
change are main \emph{universal manifestations of unreduced
dynamic complexity} and give rise to any other observed feature of
world structure and dynamics: realisation change ``produces'' a
\emph{physically real} space element $\Delta x = \Delta _r \eta
_i^r$ and time increment $\Delta t = {1 \mathord{\left/
 {\vphantom {1 \nu }} \right. \kern-\nulldelimiterspace} \nu }
 = {{\Delta x} \mathord{\left/ {\vphantom {{\Delta x} c}} \right.
 \kern-\nulldelimiterspace} c}$, where $\nu$ is the frequency
(expressing intensity) of realisation change process and $c$ is
the velocity of perturbation propagation in the system. Since the
simplest independent combination of space and time is given by
action, $\mathcal A$, we obtain the \emph{extended interpretation
of action} as a universal \emph{complexity measure} accounting for
the \emph{essentially nonlinear} realisation change process: each
realisation change cycle is described by an action increment
$\Delta {\mathcal A} =  - E\Delta t + p\Delta x$, where
coefficients $E$ and $p$ are identified as energy and momentum.

Since in the case of quantum beat dynamics one deals with the most
fundamental world structure emergence, it is impossible to observe
and measure the continuous, detailed course of the process.
Observed effects and measurements start from the complete
realisation change (reduction-extension) cycle of quantum beat,
providing the basis and realistic explanation for
\emph{quantization} (see also below), only formally introduced in
usual theory. It becomes clear also that at those \emph{lowest},
``quantum'' complexity levels $\left| {\Delta {\mathcal A}}
\right| = h$, where $h$ is Planck's constant, and one obtains the
causally complete explanation for both \emph{fixed, finite value}
of $h$ (indivisibility of the quantum beat cycle) and its
\emph{universality} for a very wide range of elementary entities
and phenomena (first level of interaction between the same two
protofields) \cite{Kir:100Quanta}, which is also taken for granted
in usual theory.

For the field-particle at rest ($p = 0$) one obtains $\Delta
{\mathcal A} =  - E_0 \Delta t$, where $E_0$ is the rest energy,
and thus
\begin{equation}\label{Eq:RestEnergy}
E_0  =  - \frac{{\Delta {\mathcal A}}}{{\Delta t}} =
\frac{h}{{\tau _0 }} = h\nu _0 \ ,
\end{equation}
where $\Delta {\mathcal A} =  - h$ (since $E_0 , \Delta t > 0$),
$\Delta t = \tau _0 $ is the \emph{emerging} ``quantum of time''
equal to a time period, $\tau _0 $, of \emph{essentially
nonlinear}, \emph{spatially chaotic} quantum beat process, and
$\nu _0 \equiv {1 \mathord{\left/ {\vphantom {1 {\tau _0 }}}
\right. \kern-\nulldelimiterspace} {\tau _0 }}$ is its frequency
($\nu _0 \sim 10^{20} \ {\rm Hz}$ for the electron). One obtains
thus the elementary, natural \emph{clock of the world} driven by
the autonomous \emph{complex-dynamical} mechanism of quantum beat
and giving rise to the most fundamental level of time
\cite{Kir:USciCom}. In agreement with the above complexity
definition, system \emph{energy} thus specified represents a
(differential) \emph{measure of dynamic complexity} (its integral
measure is given by action ${\mathcal A}$).

It is not difficult to see also that the \emph{spatially chaotic}
distribution of consecutive reduction centres of a quantum beat
process within the elementary field-particle provides it with the
universally defined \emph{property of mass} remaining otherwise
ambiguous in any unitary theory. Indeed, the quantum beat process
can be considered as \emph{chaotic wandering} of squeezed,
corpuscular state of the field-particle, also called \emph{virtual
soliton} \cite{Kir:USciCom,Kir:DoubleSol:1,Kir:DoubleSol:2,
Kir:ComDynGrav,Kir:Cosmo}, within its wavefunction, where the
basic property of \emph{inertia} emerges in the form of
\emph{resistance to change} of that chaotic motion, \emph{already
existing} in any (massive) particle, in agreement with the
heuristic concept of ``hidden thermodynamics'' of a single
particle introduced by Louis de Broglie \cite{deBroglie:1964}.
Therefore the field-particle rest energy, $E_0$, possesses,
together with the underlying quantum beat process, the intrinsic
property of inertia and should be proportional to the rest mass,
$m_0$, expressing that property, $E_0  = m_0 c^2 $ (where $c^2$ is
a coefficient to be specified later), which leads to another form
of eq.~(\ref{Eq:RestEnergy}),
\begin{equation}\label{Eq:RestMass}
m_0 c^2  = h\nu _0 \ ,
\end{equation}
proposed by de Broglie within physical derivation of the famous
expression for the particle wavelength \cite{deBroglie:These} (now
known as ``de Broglie wavelength'', see also
\cite{Kir:75MatWave}). We specify thus the \emph{dynamic origin}
of de Broglie's ``periodic phenomenon'' (quantum beat), attached
to a particle-``mobile'' that includes a localised entity (our
virtual soliton), and the intrinsic link between them,
eq.~(\ref{Eq:RestMass}).

Now, if the field-particle is set in motion, its energy-complexity
grows over the minimum of rest energy corresponding to the
homogeneous distribution of realisation probabilities (see
eq.~(\ref{Eq:ElemProbab})), while its action-complexity acquires a
coordinate dependence characterising inhomogeneous distribution of
realisation probabilities of the globally moving quantum beat
process, so that eq.~(\ref{Eq:RestEnergy}) for the particle at
rest is replaced by expression relating the total and partial
action derivatives:
\[
\frac{\Delta {\mathcal A}}{\Delta t} = \frac{\Delta {\mathcal
A}}{\Delta t} \left| _{x = {\rm const}} \right. + \frac{\Delta
{\mathcal A}}{\Delta x} \left| _{t = {\rm const}} \right.
\frac{\Delta x}{\Delta t} \ ,
\]
or
\begin{equation}\label{Eq:MotionEnergy}
E =  - \frac{{\Delta {\mathcal A}}}{{\Delta t}} + \frac{{\Delta
{\mathcal A}}}{\lambda }\frac{{\Delta x}}{{\Delta t}} =
\frac{h}{{\mathcal T}} + \frac{h}{\lambda }\thinspace v =
h{\mathcal N} + pv \ ,
\end{equation}
where
\begin{equation}\label{Eq:ParticleEnergy}
E =  - \frac{{\Delta {\mathcal A}}}{{\Delta t}}\left| {_{x  =
{\rm{const}}} } \right. = \frac{h}{\tau } = h\nu
\end{equation}
is the total system \emph{energy},
\begin{equation}\label{Eq:ParticleMomentum}
p = \frac{{\Delta {\mathcal A}}}{{\Delta x}}\left| {_{t =
{\rm{const}}} } \right. = \frac{\left| {\Delta {\mathcal A}}
\right|}{\lambda } = \frac{h}{\lambda }
\end{equation}
is its universally defined \emph{momentum},
\[
v = \frac{{\Delta x}}{{\Delta t}} \equiv \frac{\mit
\Lambda}{{\mathcal T}}
\]
is the \emph{global motion velocity}, $\lambda  \equiv \left(
{\Delta x} \right)\left| _{t = {\rm{const}}} \right.$ is the
``quantum of space'' \emph{emerging} from the field-particle
motion, $\Delta t = {\mathcal T}$ is the ``total'' period of
nonlinear quantum beat of the field-particle in the state of
motion with complexity-energy $E$ (${\mathcal N} = {1
\mathord{\left/ {\vphantom {1 {\mathcal T}}} \right.
\kern-\nulldelimiterspace} {\mathcal T}}$ is the corresponding
quantum-beat frequency), $\Delta x = {\mit \Lambda} $ is the
``total'' quantum of space, while $\tau  \equiv \left( {\Delta t}
\right)\left| _{x = {\rm{const}}} \right.$ is the quantum-beat
period measured at a fixed space point.

The wave field inhomogeneity induced by global field-particle
motion and given by eq.~(\ref{Eq:ParticleMomentum}) is none other
than the particle de Broglie wavelength $\lambda \equiv \lambda
_{\rm{B}} $, now understood thus as a result of
\emph{complex-dynamical (multivalued and probabilistic) structure
formation process}, at this complexity sublevel
\cite{Kir:75MatWave,Kir:100Quanta,Kir:75Wavefunc,
Kir:QuMach,Kir:USciCom,Kir:DoubleSol:1,Kir:DoubleSol:2,
Kir:ComDynGrav,Kir:Cosmo}. That structure emerges as averaged,
regular part of globally moving quantum beat, described by the
second summand in the particle energy partition of
eq.~(\ref{Eq:MotionEnergy}), whereas the first summand accounts
for the totally irregular (homogeneous in average) virtual soliton
wandering ``around'' that average tendency. At the same time,
every single jump of the virtual soliton preserves its dynamically
probabilistic character, so that the \emph{whole} content of the
total energy possesses the property of inertia, which provides
causal explanation and intrinsic unification of respective
``relativistic'' and quantum properties.

Since the two tendencies in the quantum beat dynamics of a
globally moving particle, the regular global displacement of the
wave field and totally random wandering of the virtual soliton
around it, form a \emph{single} complex-dynamical (multivalued)
process, another dynamic relation, properly reflecting that
essential link, should be added to the structure formation
expression of
eqs.~(\ref{Eq:MotionEnergy})--(\ref{Eq:ParticleMomentum}). If we
introduce the velocity of light $c$ as the velocity of
\emph{physically real perturbation propagation} through the e/m
protofield coupled to the gravitational protofield (as opposed to
the \emph{postulated} formal limitation of the abstract
``principle of relativity'' in standard theory), then it becomes
clear that during a time period, $\tau _1  = {\lambda
\mathord{\left/ {\vphantom {\lambda  c}} \right.
\kern-\nulldelimiterspace} c}$, of global wave field advance of
$\lambda  \equiv \lambda _{\rm{B}} $, the virtual soliton should
perform, in average, $n_1 = {c \mathord{\left/
 {\vphantom {c v}} \right. \kern-\nulldelimiterspace} v}$ irregular
jumps around it, which explains why our massive field-particle,
being an e/m protofield perturbation, moves not with the velocity
of light, but with the velocity $v < c$ (we thus \emph{derive}
this ``postulate'' of standard relativity from a physically
transparent picture of \emph{multivalued interaction} dynamics).
Since every jump duration is $\tau$, we have $n_1 \tau  = \tau _1
$, or $\lambda  = V_{{\rm{ph}}} \tau $, where $V_{{\rm{ph}}}  =
{{c^2 } \mathord{\left/ {\vphantom {{c^2 } v}} \right.
\kern-\nulldelimiterspace} v}$ is the fictitious, apparently
superluminal ``phase velocity'' of ``matter wave'' propagation,
appearing if one does not take into account the irregular,
``multivalued'' part of the field-particle dynamics
\cite{deBroglie:These}. Writing the obtained relation as ${1
\mathord{\left/ {\vphantom {1 {\lambda  = \left( {{1
\mathord{\left/ {\vphantom {1 \tau }} \right.
\kern-\nulldelimiterspace} \tau }} \right)}}} \right.
\kern-\nulldelimiterspace} {\lambda  = \left( {{1 \mathord{\left/
{\vphantom {1 \tau }} \right. \kern-\nulldelimiterspace} \tau }}
\right)}}\left( {{v \mathord{\left/ {\vphantom {v {c^2 }}} \right.
\kern-\nulldelimiterspace} {c^2 }}} \right)$, multiplying it by
$h$, and using the energy and momentum definitions,
eqs.~(\ref{Eq:ParticleEnergy}),\,(\ref{Eq:ParticleMomentum}), one
gets the famous ``relativistic'' \emph{dispersion relation}
between momentum and energy
\begin{equation}\label{Eq:DispRelation}
p = E \thinspace \frac{v}{{c^2}} = mv \ ,
\end{equation}
where $m = E / c ^2$, now by \emph{rigorously derived} definition,
in which $c$ is the above \emph{physically based} speed of light.

We obtain thus the causally complete, \emph{quantum}
interpretation of \emph{classical relativistic relation} between
energy and mass, specifying our previous assumption in
eq.~(\ref{Eq:RestMass}) and derived from the underlying complex,
multivalued and \emph{dynamically quantized} interaction dynamics.
It provides the necessary completion of the original ``phase
accord'' conjecture of Louis de Broglie \cite{deBroglie:These},
and we obtain the final expression for the de Broglie wavelength
by combining eqs.~(\ref{Eq:ParticleMomentum}) and
(\ref{Eq:DispRelation}):
\begin{equation}\label{Eq:DeBroglieWave}
\lambda  = \lambda _{\rm{B}}  = \frac{h}{{mv}} \ .
\end{equation}

A still larger implication of the causally derived dispersion
relation of eq.~(\ref{Eq:DispRelation}) is that it provides, after
(discrete or continuous) differentiation with respect to time, the
rigorous substantiation of \emph{relativistic} version of
\emph{Newton's laws} of ``classical'' dynamics, revealing the
irreducible role of underlying \emph{complex-dynamical
(multivalued)} structure emergence and including \emph{causally
complete} understanding of major entities of \emph{space, time,
energy, mass, and momentum} as various forms and measures of
\emph{dynamic complexity}. Due to universality of the above
interaction analysis, we can see that those laws and the
underlying dispersion relation, eq.~(\ref{Eq:DispRelation}),
remain valid at any complexity level, provided we have a
\emph{uniform} enough regime of (averaged) interaction
development, situated well outside of another complexity level
emergence (such as relativistic particle transformations).

Inserting the dispersion relation of eq.~(\ref{Eq:DispRelation})
into the energy partition of eq.~(\ref{Eq:MotionEnergy}) and using
energy definition of eq.~(\ref{Eq:ParticleEnergy}), one gets a
basic \emph{time relativity} expression revealing now its genuine,
\emph{complex-dynamic} origin:
\begin{equation}\label{Eq:TimeRel:1}
\tau  = {\mathcal T}\left( {1 - \frac{{v^2 }}{{c^2 }}} \right) \ .
\end{equation}
We see that our causally defined time goes more slowly ``within''
a moving entity (${\mathcal T} > \tau $) because that,
\emph{physically real} time is explicitly \emph{produced} by the
\emph{same} unceasing, spatially chaotic realisation change
process that gives rise to the global \emph{motion}. A part of
quantum beat energy that goes to the global motion is subtracted
from the irregular component of dynamics just accounting for
``internal'' time of a moving system. In order to get the standard
expression of time retardation in terms of time period in the
state of rest, $\tau _0 $, we use an additional relation between
motion frequencies $\nu$, ${\mathcal N}$, and $\nu _0 $,
\begin{equation}\label{Eq:RelPeriods}
{\mathcal N} \nu  = \left( {\nu _0} \right)^2 , \ \ {\rm or} \ \ \
{\mathcal T} \tau  = \left( {\tau _0} \right)^2 \ .
\end{equation}
These relations express a physically transparent law of
``conservation of the total number (frequency) of reduction
events'', which is a manifestation of \emph{complexity
conservation law}
\cite{Kir:USciCom,Kir:DoubleSol:1,Kir:DoubleSol:2,
Kir:ComDynGrav}. Using eq.~(\ref{Eq:RelPeriods}) in
eq.~(\ref{Eq:TimeRel:1}), we get the canonical expression of time
retardation effect, but now \emph{causally derived} in terms of
underlying \emph{complex} dynamics of the \emph{quantum} beat
process:
\begin{equation}\label{Eq:TimeRel:2}
{\mathcal N} = \nu _0 \sqrt {1 - \frac{v^2}{c^2}} \ \ \ {\rm or} \
\ \ {\mathcal T} = \frac{\tau _0}{\sqrt {\displaystyle {1 -
\frac{v^2 }{c^2}}} } \ .
\end{equation}
Related relativistic effects  (e.g. length reduction) are obtained
as straightforward consequences of these results.

We can summarise the obtained intrinsic, complex-dynamical
\emph{unification of causally complete versions of quantum and
relativistic mechanics} by combining in one expression the
complex-dynamical energy partition into regular (global) transport
and irregular (local) wandering, eq.~(\ref{Eq:MotionEnergy}), with
the causally derived dispersion relation,
eqs.~(\ref{Eq:DispRelation}),\,(\ref{Eq:DeBroglieWave}), and time
(frequency) relation to dynamics, eqs.~(\ref{Eq:TimeRel:2}):
\begin{eqnarray}\label{Eq:EnergyPartition}
\displaystyle
 E = h\nu _0 \sqrt {1 - \frac{{v^2 }}{{c^2 }}}  +
\frac{h}{{\lambda _{\rm{B}} }} \thinspace v = h\nu _0 \sqrt {1 -
\frac{{v^2 }}{{c^2}}}  + h\nu _{\rm{B}}  = \nonumber \\[+8 pt]
\displaystyle
 = m_0 c^2 \sqrt {1 - \frac{{v^2 }}{{c^2 }}} +
\frac{{m_0 v^2 }}{{\sqrt {\displaystyle {1 - \frac{{v^2 }}{{c^2
}}}} }} \ \ , \hspace {1.75 cm}
\end{eqnarray}
where $h\nu _0  = m_0 c^2 $ (eq.~(\ref{Eq:RestMass})) and we
introduce \emph{de Broglie frequency}, $\nu _{\rm{B}} $, defined
as
\begin{equation}\label{Eq:DeBroglieFrequency}
\nu _{\rm B}  = \frac{v}{{\lambda _{\rm B} }}  = \frac{{pv}}{h} =
\frac{{\nu _{{\rm B} 0} }}{{\sqrt {\displaystyle {1 - \frac{{v^2
}}{{c^2 }}}} }} = \nu \thinspace \frac{{v^2 }}{{c^2 }} \ ,
\end{equation}
\[
\nu _{{\rm{B}}0} = \frac{{m_0 v^2 }}{h}  = \nu _0 \thinspace
\frac{{v^2 }}{{c^2 }} = \frac{v}{{\lambda _{{\rm{B}}0} }} \ , \ \
\lambda _{{\rm{B}}0}  = \frac{h}{{m_0 v}} \ .
\]
Physical \emph{reality} of de Broglie wave, obtained as a result
of dynamically multivalued \emph{structure formation process} in
the system of two protofields, is confirmed by ``usual'' relation
between its wavelength, frequency, and velocity, $\lambda
_{\rm{B}} \nu _{\rm{B}}  = v$, comprising a nontrivial,
\emph{essentially nonlinear} and \emph{causally random},
alternation of \emph{dual}, undular \emph{and} corpuscular,
protofield structures \emph{permanently transformed into one
another} and \emph{thus} preserving their ``phase accord''
\cite{deBroglie:These,
deBroglie:1927,deBroglie:1952,deBroglie:1956,deBroglie:1959,deBroglie:1997,
deBroglie:1964}. It is not difficult to see from
eq.~(\ref{Eq:EnergyPartition}) \cite{Kir:DoubleSol:2,
Kir:ComDynGrav} that the quantities $\alpha _1 = {{v^2 }
\mathord{\left/ {\vphantom {{v^2 } {c^2 }}} \right.
\kern-\nulldelimiterspace} {c^2 }}$ and $\alpha _2  = 1 - \alpha
_1  = 1 - {{v^2 } \mathord{\left/ {\vphantom {{v^2 } {c^2 }}}
\right. \kern-\nulldelimiterspace} {c^2 }}$ represent
\emph{dynamically} obtained \emph{realisation probabilities} for,
respectively, global (regular) and totally random tendencies of
the moving field-particle dynamics, in full agreement with the
general definition of eq.~(\ref{Eq:RealProbab}), confirming once
more the intrinsic unity of ``quantum'' and ``relativistic''
manifestations of unreduced interaction complexity.

The first term in the sum of
eqs.~(\ref{Eq:MotionEnergy}),\,(\ref{Eq:EnergyPartition}) taken
with the opposite sign, ${{\Delta {\mathcal A}} \mathord{\left/
{\vphantom {{\Delta {\mathcal A}} {\Delta t}}} \right.
\kern-\nulldelimiterspace} {\Delta t}} = pv - E \equiv L$, is
known as system \emph{Lagrangian}. Energy partition of
eq.~(\ref{Eq:EnergyPartition}) provides thus the \emph{causally
derived} expression for the Lagrangian, together with its physical
meaning of purely random, ``thermal'' part of multivalued
(chaotic) system dynamics taken with the opposite, negative sign:
\[
L =  - h{\mathcal N} =  - h\nu _0 \sqrt {1 - \frac{{v^2 }}{{c^2
}}} = - m_0 c^2 \sqrt {1 - \frac{{v^2 }}{{c^2 }}} \ .
\]
This interpretation leads to the causally derived, realistic
extension of the ``principle of least action''
\cite{Kir:USciCom,Kir:DoubleSol:2, Kir:ComDynGrav}, anticipated by
Louis de Broglie, whereas standard relativity \emph{postulates}
that principle and mechanistically \emph{guessed}, formal
expression for the Lagrangian and action (see e.g.
\cite{Einstein,Landau:FT}), tacitly adding these and other
assumptions, essentially used in the theory, to the explicitly
announced ``principle of relativity''. The obtained negative sign
of the Lagrangian is its universal property expressing the
\emph{dynamic arrow (irrevisibility) of time} and its orientation
to \emph{growing dynamic entropy} and correspondingly decreasing
dynamical information just represented by action-complexity, so
that their sum, the total system complexity, remains unchanged
\cite{Kir:QuMach,Kir:USciCom,Kir:USymCom}.

The multivalued version of elementary particle dynamics should be
completed by explicit equation for its undular, wavefunctional
state, causally introduced above (whereas action and its
derivatives refer mainly to the localised state of virtual
soliton). That \emph{wave equation} is obtained with the help of
\emph{causal quantization} procedure
\cite{Kir:100Quanta,Kir:75Wavefunc,Kir:QuMach,Kir:USciCom,Kir:USymCom,Kir:DoubleSol:2,
Kir:ComDynGrav,Kir:Cosmo} summarising quantum beat dynamics, which
returns to the same state of wavefunction (intermediate
realisation) in each reduction-extension cycle. It corresponds to
system complexity conservation expressed by the product of
action-complexity of corpuscular state, $\mathcal A$, and that of
the wavefunction, ${\mit \Psi}$, so that $\Delta ({\mathcal A}
{\mit \Psi}) = 0$ for each quantum beat cycle:
\begin{equation}\label{Eq:CausQuant:1}
\Delta \left( {{\mathcal A} {\mit \Psi} } \right) = {\mathcal A}
\Delta {\mit \Psi}  + {\mit \Psi} \Delta {\mathcal A} = 0
\thinspace , \ {\rm or} \ \Delta {\mathcal A} = - \hbar \thinspace
\frac{{\Delta {\mit \Psi} }}{{\mit \Psi} } \thinspace ,
\end{equation}
since the characteristic value of $\mathcal A$ is equal to $\hbar
= {h \mathord{\left/ {\vphantom {h {2 \pi }}} \right.
 \kern-\nulldelimiterspace} {2 \pi }}$. The latter quantity
is then additionally multiplied by the imaginary unit, $i$, which
does not change the physical sense of quantization and accounts
for the difference between wave and corpuscular states in wave
presentation by complex numbers:
\begin{equation}\label{Eq:CausQuant:2}
\Delta {\mathcal A} =  - i \hbar \thinspace \frac{{\Delta {\mit
\Psi} }}{\mit \Psi} \ .
\end{equation}
Causal version of differential, ``Dirac'' quantization rules is
then obtained by using the above definitions of momentum,
eq.~(\ref{Eq:ParticleMomentum}), and energy,
eq.~(\ref{Eq:ParticleEnergy}):
\begin{equation}\label{Eq:DiracQuant:Momentum}
p = \frac{\Delta {\mathcal A}}{\Delta x} =  - \frac{1}{\mit \Psi}
\thinspace i \hbar \thinspace \frac{\partial {\mit \Psi}}{\partial
x} \ , \ \ p^2 = - \frac{1}{\mit \Psi} \thinspace \hbar ^2
\frac{\partial ^2 {\mit \Psi} }{\partial x^2} \ ,
\end{equation}
\begin{equation}\label{Eq:DiracQuant:Energy}
E = - \frac{\Delta {\mathcal A}}{\Delta t} =  \frac{1}{\mit \Psi}
\thinspace i \hbar \thinspace \frac{\partial {\mit \Psi}}{\partial
t} \ , \ \ E^2 = - \frac{1}{\mit \Psi} \thinspace \hbar ^2
\frac{\partial ^2 {\mit \Psi} }{\partial t^2} \ ,
\end{equation}
where quantization of higher powers of $p$ and $E$ properly
reflects the wave nature of $\mit \Psi$
\cite{Kir:75Wavefunc,Kir:USciCom}, and $x$ can be directly
extended to a three-dimensional vector.

Inserting now the obtained quantization expressions into a version
of eq.~(\ref{Eq:EnergyPartition}),
\begin{equation}\label{Eq:EnergyPartition'}
E = m_0 c^2 \sqrt {1 - \frac{{v^2 }}{{c^2 }}}  + \frac{{p^2 }}{m}
\ \ \ \ {\rm or} \ \ \ mE = m_0 c^2  + p^2 \ ,
\end{equation}
we get the desired equation for $\mit \Psi$, equivalent to the
simplest form of both Dirac and Klein-Gordon equations:
\begin{equation}\label{Eq:Dirac}
i \hbar m \thinspace \frac{{\partial {\mit \Psi} }}{{\partial t}}
+ \hbar ^2 \frac{ {\partial ^2 {\mit \Psi} }}{{\partial x^2 }} -
m_0^2 c^2 {\mit \Psi} = 0 \ ,
\end{equation}
\begin{equation}\label{Eq:Klein-Gordon}
\frac{\partial ^2 {\mit \Psi}}{\partial t^2} -  c^2 \frac{\partial
^2 {\mit \Psi}}{\partial x^2} + \omega _0^2 {\mit \Psi} = 0 \ ,
\end{equation}
where $\omega _0  \equiv m_0 c^2 / \hbar = 2 {\pi} \nu _0$ is the
``circular'' frequency of quantum beat pulsation at rest (see
eq.~(\ref{Eq:RestMass})) accounting for the spin vorticity twirl
\cite{Kir:USciCom,Kir:DoubleSol:1,Kir:DoubleSol:2, Kir:ComDynGrav}
(see also below). More elaborated forms of wave equation, taking
into account particle interactions, can be obtained within the
same causal quantization procedure \cite{Kir:USciCom}, endowing
formally identical usual versions with a realistic,
complex-dynamic substantiation. In the nonrelativistic limit they
are reduced to the Schr\"{o}dinger equation, but provided with the
causally complete interpretation.

One can also obtain Schr\"{o}dinger equation by using causal
quantization rules,
eqs.~(\ref{Eq:DiracQuant:Momentum})--(\ref{Eq:DiracQuant:Energy}),
in the nonrelativistic limit of energy-momentum relation of
eq.~(\ref{Eq:EnergyPartition'}) for particle in external
potential, $V\left( {x,t} \right)$ (which causally appears as a
dynamically ``folded'', ``potential'' form of complexity, or
``dynamic information''
\cite{Kir:QuMach,Kir:USciCom,Kir:USymCom,Kir:Cosmo}):
\begin{eqnarray}
E =  \frac{{p^2 }}{{2m_0 }} + V\left( {x,t} \right) \to \hspace
{1.2 cm} \nonumber \\[+6 pt]
 \hspace {0.7 cm} \to \ \ i \hbar \thinspace
\frac{{\partial {\mit \Psi} }}{{\partial t}} =   - \frac{{\hbar ^2
}}{{2m_0 }} \ \frac{{\partial ^2 {\mit \Psi} }}{{\partial x^2 }} +
V\left( {x,t} \right){\mit \Psi} \left( {x,t} \right) \thinspace .
\label{Eq:Schroedinger}
\end{eqnarray}
It can be shown \cite{Kir:USciCom} that for binding potentials
eq.~(\ref{Eq:Schroedinger}) can be satisfied only for discrete
configurations of ${\mit \Psi} \left( x \right)$ characterised by
integer numbers of the \emph{same action-complexity quantum}, $h$,
that describes one quantum-beat cycle, or ``system realisation
change'', which explains the famous ``energy-level discreteness''
by \emph{universal complex-dynamic} discreteness (or
\emph{dualistic} quantization) of \emph{unreduced} interaction
process. It means also that quantum-mechanical ``linear
superposition'' of eigenfunctions and respective eigenvalue
probabilities, including the special case of ``quantum
entanglement'' for a many-body system, reflects \emph{multivalued
dynamics} of the underlying \emph{interaction}, where the system
performs \emph{unceasing} ``quantum jumps'' (reduction-extension
cycles) between eigen-states (see
eqs.~(\ref{Eq:ProbabSum}),\,(\ref{Eq:EP-Full})--(\ref{Eq:StFunc-Full}))
with \emph{dynamically determined} probabilities
(eqs.~(\ref{Eq:ElemProbab})--(\ref{Eq:RealProbab})).
\emph{Quasi}-linearity of wavefunction behaviour is due to
\emph{transiently} weak, perturbative interaction character
\emph{only} within that particular, intermediate realisation of
the wavefunction (see above), whereas measured eigenvalue
emergence from that realisation, hidden within ``inexplicable''
standard postulates, is due to \emph{essentially nonlinear} and
\emph{physically real} creation of respective regular
realisations.

Similar to mass, \emph{every intrinsic property} of elementary
field-particle can be \emph{causally} explained as a manifestation
of \emph{physically real}, but \emph{irreducibly complex},
essentially nonlinear quantum-beat dynamics. Thus, the property of
\emph{spin} results from ``shear instability'' of collapsing
protofield dynamics, leading to highly nonlinear vorticity
emergence, similar to a weakly nonlinear case of a liquid escaping
through a small hole from a basin under the influence of gravity
\cite{Kir:USciCom} (and similar to a water twirl, the direction of
spin vorticity, determining particle \emph{helicity}, may be due
to a weak rotational, or ``CP'', asymmetry of the driving
protofield interaction). The same rest energy of a particle,
eq.~(\ref{Eq:RestEnergy}), can then be presented, for example for
a spin-1/2 particle like the electron, as $E_0  = h\nu _0 = \hbar
\omega _0  = 2s\omega _0 $, where $s = {\hbar \mathord{\left/
{\vphantom {\hbar  2}} \right. \kern-\nulldelimiterspace} 2}$ is
the spin angular momentum for each phase of reduction or extension
of the quantum beat dynamics. Particle \emph{magnetic moment} and
\emph{magnetic field} originate from the same quantum beat
vorticity (in the extension phase) \cite{Kir:USciCom}.

The property of \emph{electric charge} expresses the
\emph{long-range} interaction between individual quantum beat
processes through the common e/m medium. In view of the well-known
proportionality relation between the squared elementary charge $e$
and Planck's constant $\hbar$, $\alpha \hbar  = {{e^2 }
\mathord{\left/ {\vphantom {{e^2 } c}} \right.
\kern-\nulldelimiterspace} c}$ (where $\alpha  \approx {1
\mathord{\left/ {\vphantom {1 {137}}} \right.
\kern-\nulldelimiterspace} {137}}$ is the famous fine structure
constant), it becomes clear that $e$ expresses the same
\emph{dynamically discrete} structure of protofield interaction
complexity as $h$ (corresponding to one quantum beat cycle), but
now with respect to \emph{phase-sensitive} e/m interactions
between different quantum beat processes. Universal
\emph{quantization} of electric charge in units of $e$ acquires
thus a \emph{dynamic}, causal origin, directly related to causal
quantization of motion. The number (two) and ``opposite'' nature
of existing kinds of charge are explained by \emph{universal}
complex-dynamical phase synchronisation of individual quantum beat
processes up to the opposite phase: quantum beat processes with
opposite phases appear as unlike, ``opposite'' charges,
\emph{dynamically} attracted to each other \cite{Kir:USciCom}. In
accord with our physically real origin of time, the same quantum
beat synchronisation accounts for \emph{universality of time
flow}, also taken for granted in usual theory.

The standard relation between $e$ and $h$ can be presented in a
form providing better insight into the underlying quantum beat
dynamics:
\begin{equation}\label{Eq:FineStrPlanck}
m_0 c^2  = \frac{ 2 \pi }{\alpha } \thinspace \frac{{e^2
}}{{\lambda _{\rm{C}} }} = N_\Re ^e \thinspace \frac{{e^2
}}{\mathchar'26\mkern-10mu\lambda _{\rm C}} \ , \ N_\Re ^e =
\frac{1}{\alpha} \ , \ \mathchar'26\mkern-10mu\lambda _{\rm C} =
\frac{\lambda _{\rm C}}{2 \pi} \ ,
\end{equation}
where $\lambda _{\rm{C}}  = {h \mathord{\left/ {\vphantom {h {m_0
c}}} \right. \kern-\nulldelimiterspace} {m_0 c}}$ is the Compton
wavelength of the elementary field-particle with the rest mass
$m_0$ and electric charge $e$ (actually represented by the
electron or positron). The electron rest energy can be considered
thus as a sum of $N_\Re ^e   = {{1} \mathord{\left/ {\vphantom
{{1} \alpha }} \right. \kern-\nulldelimiterspace} \alpha }$ e/m
interactions between two elementary charges at a distance of
$\mathchar'26\mkern-10mu\lambda _{\rm C}$. But since multivalued
quantum beat dynamics consists of virtual soliton jumps between
its realisations (reduction centres), we can consistently
interpret the above relation by assuming that
$\mathchar'26\mkern-10mu\lambda _{\rm C}$ expresses the length of
elementary particle jump between its two consecutive localised
realisations and $N_\Re ^e   = {{1} \mathord{\left/ {\vphantom
{{1} \alpha }} \right. \kern-\nulldelimiterspace} \alpha } \approx
137$ gives (up to its fractional part) the total number of system
realisations, i. e. the number of possible reduction centre
positions for the electron or, equivalently, the number of the
constituent \emph{virtual photons} in its \emph{extended} phase.
We obtain thus the new, \emph{intrinsic} and
\emph{complex-dynamical} interpretation of both Compton length and
fine structure constant. It implies, in particular, that the fine
structure constant, $\alpha = {{1} \mathord{\left/ {\vphantom {{1}
{N_\Re ^e}}} \right. \kern-\nulldelimiterspace} {N_\Re ^e}}$,
coincides with the elementary realisation probability, $\alpha _r
= {1 \mathord{\left/ {\vphantom {1 {N_\Re }}} \right.
\kern-\nulldelimiterspace} {N_\Re }}$, defined by
eq.~(\ref{Eq:ElemProbab}).

The obtained expression for the Compton length corresponds to the
de Broglie wavelength, eq.~(\ref{Eq:DeBroglieWave}), for a
particle \emph{simultaneously} moving with the speed of light ($v
= c$) \emph{and} remaining \emph{globally} at rest ($m = m_0 $).
That ``impossible'' combination of properties is practically
realised, however, just for virtual soliton jumps \emph{within}
the electron remaining globally at rest but having
\emph{individual} jump velocity equal to $c$.

Preservation of spin orientation in the absence of external
influences implies that all the next possible $N_\Re ^e$
realisation centres for the electron are situated on a
two-dimensional circle with a radius of
$\mathchar'26\mkern-10mu\lambda _{\rm C}$ and length equal to  $2
\pi \mathchar'26\mkern-10mu\lambda _{\rm C} = \lambda _{\rm{C}} $.
Since the system of realisations is complete, they should fill in
the circle length without remaining free space, so that the size
of each localised realisation (or virtual soliton) is $D_e  =
{{\lambda _{\rm{C}} } \mathord{\left/ {\vphantom {{\lambda
_{\rm{C}} } {N_\Re ^e}}} \right. \kern-\nulldelimiterspace} {N_\Re
^e}} = {{2\pi e^2 } \mathord{\left/ {\vphantom {{2\pi e^2 } {m_0
c^2 }}} \right.
 \kern-\nulldelimiterspace} {m_0 c^2 }} = 2\pi r_e  = \pi d_e $
(or $\lambda _{\rm{C}}  = N_\Re ^e  D_e$), where $r_{e} = {{e^2 }
\mathord{\left/ {\vphantom {{e^2 } {m_0 c^2 }}} \right.
\kern-\nulldelimiterspace} {m_0 c^2 }}$ is the well-known
``classical electron radius'' (and $d_{e}  = 2r_{e}$ the
corresponding diameter), obtained by expression of the electron
rest energy as ``purely e/m (Coulomb) self-interaction''. This
relation demonstrates the degree of nonlinear dynamical squeeze of
e/m protofield in the virtual soliton state of the electron, which
is of the order of $\left( 2\mathchar'26\mkern-10mu\lambda _{\rm
C} / D_e \right)^3 = \left( {{{N_\Re ^e} \mathord{\left/
{\vphantom {{N_\Re ^e} { \pi }}} \right.
\kern-\nulldelimiterspace} { \pi }}} \right)^3  = \left( \pi
\alpha \right)^{-3} \approx 0,83 \times 10^5 \sim 10^5 $, in terms
of three-dimensional volume contraction. On the other hand, it
provides a causally specified extension of usual, rather vague
$\alpha$ interpretation as e/m interaction ``strength'': $N_\Re ^e
= {{1} \mathord{\left/ {\vphantom {{1} \alpha }} \right.
\kern-\nulldelimiterspace} \alpha }$ characterises the
\emph{width} of EP potential well of the protofield interaction
realisation for the electron, which is inversely proportional to
its \emph{depth} (and thus $\alpha$ is \emph{proportional} to the
depth) because the product of width by depth remains constant.
Indeed, one can write $\hbar = \mathchar'26\mkern-10mu\lambda
_{\rm C} p_e$, where $p_e = m_0 c = {{E_0 } \mathord{\left/
{\vphantom {{E_0 } c}} \right. \kern-\nulldelimiterspace} c} =
{{e^2 } \mathord{\left/ {\vphantom {{e^2 } {r_e c}}} \right.
\kern-\nulldelimiterspace} {r_e c}}$ is the expression of the
protofield EP depth in terms of momentum ($E_0$ is the EP depth
value in terms of energy) and $\mathchar'26\mkern-10mu\lambda
_{\rm C} = N_\Re ^e  r_e = {{r_e} \mathord{\left/ {\vphantom
{{r_e} \alpha }} \right. \kern-\nulldelimiterspace} \alpha }$ is
the corresponding EP well width, so that $\alpha  = {{p_e r_e}
\mathord{\left/ {\vphantom {{p_e r_e} \hbar }} \right.
\kern-\nulldelimiterspace} \hbar } = {{E_0 r_e} \mathord{\left/
{\vphantom {{E_0 r_e} {c\hbar }}} \right.
\kern-\nulldelimiterspace} {c\hbar }} = {{e^2 } \mathord{\left/
{\vphantom {{e^2 } {c\hbar }}} \right. \kern-\nulldelimiterspace}
{c\hbar }}$ and $\hbar = N_\Re ^e \left( {{{E_0 r_e }
\mathord{\left/ {\vphantom {{E_0 r_e } c}} \right.
\kern-\nulldelimiterspace} c}} \right) = {{N_\Re ^e  e^2 }
\mathord{\left/ {\vphantom {{N_\Re ^e e^2 } c}} \right.
\kern-\nulldelimiterspace} c}$.

One obtains thus a realistic \emph{interpretation of Planck's
constant} as the ``volume'' of EP well for the protofield
interaction, i.\,e. the product of EP well depth and width, for
\emph{any} elementary particle (emerging in the complex-dynamical
interaction development), which explains the unlimited
\emph{universality} of $h$ and the origin of \emph{different
particle species} as different protofield interaction realisations
with varying EP depths and widths, but permanent product of the
two \cite{Kir:100Quanta}. While the electron emerges as a massive
field-particle with the widest and most shallow EP well (largest
$N_\Re$ and smallest $\alpha$, $m_0$, and $e^2$), the heaviest,
``strongly'' interacting particles correspond to the most narrow
EP well (effective $N_\Re  , \alpha  \sim 1$) and its biggest
depths (maximum $m_0$ and ``self-interaction'' values). Two higher
``generations'' of elementary particles emerge as ``excited'' and
\emph{therefore} more chaotic (less stable) protofield EP
realisations with increased amplitude (and correspondingly
decreased widths), where the phenomenon of dynamic multivaluedness
explains the very fact of \emph{multiple} generations existence,
remaining otherwise ``mysterious'' in the unitary theory
framework, as well as the origin and basic features of particle
spectrum in general (see also ref.~\cite{Baten} for possible
further elaboration of detailed particle properties in a realistic
framework inspired by quantum field mechanics).

Since (extended) action represents, as we have seen, a universal
measure of unreduced dynamic complexity, it is clear why the
magnitude of protofield interaction is fixed as an action
constant: the latter determines actually the quantity of
``latent'', ``potential'' complexity, or ``dynamic information'',
or ``protofield separation work'', that must be ``invested'' into
the protofield system (in the form of attracting protofield
separation) in order that it can produce the observed world
structure and dynamics, in agreement with the universal symmetry
(conservation) of complexity
\cite{Kir:QuMach,Kir:USciCom,Kir:USymCom}.

The origin, number, and properties of \emph{fundamental
interaction forces} between individual elementary particles are
also causally and naturally derived in quantum field mechanics,
together with their \emph{intrinsic, dynamic unification}. Two
most universal, long-range forces, the e/m and gravitational
interactions emerge as inevitable interactions between individual
quantum beat processes transmitted through, respectively, the
common e/m and gravitational protofields starting from protofield
deformations (tensions) that arise around every quantum beat
pulsation. The detailed e/m interaction mechanism involves
causally extended \emph{exchange of real ``virtual'' (transient)
photons} between different quantum beats, where photons are
obtained as massless (but very slightly dissipative),
\emph{dynamically} quantized, and shallow-EP excitations of e/m
protofield stabilised by its interaction (effectively weak in
\emph{this} case) with the gravitational protofield.

The absence of (long-range) gravitational repulsion shows that the
gravitational protofield is \emph{qualitatively} different, by its
physical nature, from the e/m protofield and resembles rather a
dense, dissipative medium, where all waves (excitations)
transmitting interactions quickly decay as such, together with
their phases that remain essential for the case of highly
``elastic'' e/m protofield (it is easy to see that such kind of
difference between the ``materials'' of the two protofields
\emph{should} exist in any case for efficient structure formation
during their interaction, which devaluates all resource-consuming
efforts of conventional science around long-range gravitational
wave detection and usual quantization of gravity). The general,
averaged ``tension'' is, however, always transmitted through the
gravitational medium, giving rise to \emph{universal} gravitation.

Two other, short-range interaction forces, ``strong'' and ``weak''
interactions, are conveniently explained as short-range
interaction forces between discrete structure elements of
gravitational and e/m protofields, respectively. It leads one to
the conclusion about the physical origin of gravitational medium
that appears to be a sort of ``quark matter'', or ``condensate'',
which provides the \emph{natural, physical unification between
gravity and strong interaction} as between long- and short-range
forces transmitted by the \emph{same} medium, by analogy to the
known ``electro-weak'' unification, which is only \emph{formally}
established in conventional theory, but now can be \emph{causally
understood} as \emph{physically unified} long- and short-range
interactions within the e/m protofield.

The \emph{number of fundamental forces} (four), their
\emph{origin} and \emph{properties} appear now \emph{exactly as
they should be} for interaction between two physically real
protofields, where the forces can be ``symmetrically'' grouped
into two couples by two principles, according to either their
``protofield of origin'' (electro-weak and gravi-strong
interaction couples), or range of transmission (universality of
perception of e/m-gravitational and weak-strong interaction
couples). Moreover, \emph{all the four forces} are naturally,
\emph{dynamically unified} within the quantum beat process
(actually in its realisation for strongly interacting particles,
hadrons), where the direct, complete unification is attained in
the phase of maximum dynamical squeeze (for the heaviest
particles). One can compare this physically transparent and
totally consistent kind of unification (suggesting, of course,
further elaboration of details) with purely abstract and
inevitably failing ``unification'' efforts in conventional,
unitary theory.

It is clear from the above picture that \emph{each} emerging
interaction between quantum beat processes has a \emph{dynamically
discrete}, naturally \emph{quantized} origin, which is especially
important for the case of \emph{gravity} that escapes any
conventional quantization attempts. Moreover, it becomes clear why
the \emph{geometric} ``model'' and approach of Einsteinian general
relativity \emph{cannot} provide a \emph{consistent} description
of the \emph{intrinsically chaotic} (multivalued) quantum beat
dynamics, representing the universal, physically real mechanism of
\emph{any} ``quantization'', while a purely \emph{formal} mixture
between \emph{quantized, tangible space} and \emph{irreversibly
flowing, immaterial time}, attempted in the standard relativity
framework, leads to deep contradictions on any scale.

By contrast, \emph{intrinsically quantized} gravity of quantum
field mechanics naturally leads to major observed effects of
``relativistic'' gravitation on a \emph{macroscopic} scale, which
do not originate in abstract, formally imposed postulates and
``principles'', but \emph{causally emerge} from the \emph{same}
unreduced, \emph{complex} dynamics of \emph{microscopic}
interaction between the two protofields that gives particles
themselves, their causally explained ``quantum'' properties and
effects of ``special'' relativity, thus providing the
\emph{intrinsically unified} (and essentially
\emph{complex-dynamical}) framework for the \emph{whole}
fundamental physics, helplessly missing in standard, unitary
theories \cite{Kir:USciCom,Kir:DoubleSol:2,
Kir:ComDynGrav,Kir:Cosmo}. In particular, the key effect of
\emph{time retardation} in gravitational field is causally derived
in our approach as quantum beat frequency dependence on the local
tension of the gravitational protofield created by other
field-particles and expressed as the gravitational field
``potential'':
\begin{equation}\label{Eq:GravTimeRetard}
h\nu _0 ( x ) = m_0 c^2 \sqrt {g_{00} ( x )} \ ,
\end{equation}
where $\nu _0 ( x )$ is the local quantum beat frequency, while
``metric'' $g_{00} ( x )$ reflects in reality a relation to the
gravitational protofield tension (or ``potential'') taking the
form, for the case of weak fields, $g_{00} \left( x \right) = 1 +
2{{\phi _g \left( x \right)} \mathord{\left/ {\vphantom {{\phi _g
\left( x \right)} {c^2 }}} \right. \kern-\nulldelimiterspace} {c^2
}}$, where $\phi _g ( x )$ is the classical gravitational field
potential \cite{Landau:FT}. Since $\nu _0 ( x )$ determines the
\emph{causally derived} ``time flow'' (see above) and $\phi _g ( x
)$ has a negative sign ($g_{00} \left( x \right) < 1$)
corresponding to gravitational attraction,
eq.~(\ref{Eq:GravTimeRetard}) substantiates a \emph{causal,
dynamical} version of ``relativistic time retardation'' in a
gravity field. It is evident that ``paradigmatic'' light-bending
effect of gravity is easily derived in a similar causal way as a
physical ``refraction'', rather than ``geometric'', phenomenon.

Another important implication of the causal origin of gravity is
\emph{renormalisation of Planckian units} and solution of the
\emph{``hierarchy'' problem} in particle \emph{mass spectrum}
\cite{Kir:USciCom,Kir:ComDynGrav,Kir:Cosmo}, among many other
modifications involved with those units. The problem arises from
the fact that the absent real, dynamic origin of phenomena is
replaced, in unitary theory, by a formal, mechanistic play with
parameters leading, in particular, to confusion between the
ordinary gravitational constant, accounting for \emph{indirect}
gravitational interaction \emph{between different particles}, and
the magnitude of the underlying \emph{direct} protofield
interaction \emph{within} each particle. Since the direct
protofield interaction within a particle is much stronger than
usual, indirect gravitational interaction between particles, the
ordinary gravitational constant in the Planckian unit definition
should be replaced by a much higher value ($10 ^ {33} - 10 ^ {34}$
times higher), which leads just to the ``right'' values of
Planckian units, where the Planckian mass value coincides (by
order of magnitude) with that of the heaviest known particles
($\sim 100 \ {\rm GeV}$), and time and length units also give
respective observed values
\cite{Kir:USciCom,Kir:ComDynGrav,Kir:Cosmo}. This result, obtained
without contradictory introduction of artificial entities, such as
``invisible dimensions'', reveals the true, physical meaning of
Planckian units: they describe the parameters of ``extreme'',
highest-magnitude EP and quantum beat (particle) realisation for
the given protofield interaction (fixed $h$ value, see above).

That intrinsic causality persists for all world structures that
emerge unevenly in the interacting protofield system as ``levels''
of complexity-entropy progressively growing at the expense of
decreasing complexity-action, or ``dynamic information''
\cite{Kir:QuMach,Kir:USciCom,Kir:USymCom,Kir:SelfOrg,Kir:Fractal:1,
Kir:Fractal:2,Kir:Conscious,Kir:ComDynGrav,Kir:Cosmo}. Thus,
\emph{classical}, permanently localised behaviour emerges as the
\emph{next complexity level} of \emph{elementary bound} (closed)
system, such as atom, due to \emph{dynamic randomness} of each
constituent quantum beat process: several \emph{bound} virtual
solitons cannot perform an extended bound walk with a
non-negligible probability just because of \emph{independent}
chaoticity of \emph{individual} quantum beats. It is the simplest
manifestation of a general regime of ``multivalued
self-organisation'', which complements the opposite case of
``uniform chaos'' (exemplified by the isolated quantum beat
process) and where, contrary to usual self-organisation, the true
chaos (complexity-entropy) \emph{persists} and \emph{increases},
but also becomes ``confined'' within a ``regular'' external shape
\cite{Kir:QuMach,Kir:USciCom,Kir:SelfOrg}.

In a similar way, the unreduced, dynamically multivalued analysis
of nonbinding interactions leads to the causally complete theory
of \emph{true quantum chaos}
\cite{Kir:QuMach,Kir:USciCom,Kir:QuChaos,Kir:Channel} and
\emph{quantum measurement} \cite{Kir:USciCom,Kir:QuMeasure} (both
of them involving genuine randomness of the same origin), which
extends essentially usual theory suffering just from the absence
of intrinsic randomness and dynamic entanglement in their unitary
basis. Note also causal solution of the famous \emph{quantum
entanglement} ``mystery'' in terms of dynamic, interaction-driven
entanglement (see discussions after eqs.~(\ref{Eq:ExpValue}) and
(\ref{Eq:Schroedinger})) and the ensuing clarification of
\emph{quantum many-body problems}
\cite{Kir:75Wavefunc,Kir:QuMach,Kir:USciCom}.

\section{Practical consequences and experimental\\ confirmation
of quantum field mechanics}\label{Sec:ExpConf}
Having described, in the previous section, the fundamental
framework and the ensuing structure of reality, we can now present
a summary of results of more immediate practical importance
(already mentioned above) and also those that provide experimental
support for the theory, especially vs unitary framework. Practical
implications of quantum field mechanics are described also in
\cite{Kir:75MatWave,Kir:100Quanta,Kir:75Wavefunc,
Kir:QuMach,Kir:USciCom,Kir:DoubleSol:2,
Kir:ComDynGrav,Kir:Cosmo,Kir:QuChaos,Kir:QuMeasure,Kir:Channel}
(see e.g. section 3 in \cite{Kir:100Quanta}).

(1) The first ``practical'' implication to be noted is still
closer to \emph{\textbf{unified consistency of quantum field
mechanics}}, which is \emph{inseparable} from the \emph{totality}
of observed phenomena. It is important to emphasize the difference
with respect to ``experimental verification'' criterion of unitary
science that prefers to emphasize \emph{separate} points of
\emph{quantitative} agreement often obtained by \emph{mechanistic
adjustment} of desired number of ``free'' parameters and
artificial, abstract entities, while at the same time it accepts,
and often tries to mask, its para-scientifically high and
\emph{growing} number of strong, \emph{qualitative} contradictions
and deviations from reality, which are often even ``postulated''
as allegedly ``unavoidable'', but actually desirable,
``mysteries''.

Causally complete analysis of the universal science of complexity
shows why that evident intellectual fraud is \emph{inevitable} in
the ``new'' fundamental physics: it tries to realise the
\emph{basically impossible} unification of \emph{explicit}, strong
manifestations of \emph{unreduced complexity} (dynamically
multivalued entanglement), constituting the \emph{true meaning} of
the ``new physics'' phenomena, with the dynamically single-valued,
\emph{zero-complexity} framework of its unitary ``paradigm'' (see
also section \ref{Sec:Intro}).

By contrast, the unreduced, \emph{complex-dynamical} description
of reality suggests another, \emph{causally complete} kind of
agreement with the observed world properties, in the form of
\emph{consistent system of correlations}, including \emph{explicit
derivation} of \emph{real} entities and their properties, where
\emph{qualitative} aspects play a \emph{greater} role than
separate cases of quantitative coincidence, since the former
cannot be adjusted (while the \emph{number} of entities should
remain at its \emph{provable} minimum).

Among general correlations within quantum field mechanics we can
mention the \emph{number of space dimensions} (three) that
\emph{emerge dynamically} from the same number of interaction
entities (two protofields and interaction itself) due to the
universal complexity conservation law (which is causally justified
and supported by the \emph{totality} of existing observations
\cite{Kir:USciCom}). The unceasing and spatially chaotic time
flow, naturally dualistic field-particles with universally defined
mass, energy, electric charge, spin, unified quantum and
relativistic properties without ``mysteries'', the number and
properties of dynamically unified interaction forces between
particles, intrinsic classicality origin, true quantum chaos and
dynamically probabilistic quantum measurement continue the list of
\emph{explicitly obtained} entities and \emph{unified
correlations} of quantum field mechanics (section
\ref{Sec:UniDynEmerge}), which minimises the probability of
artificially ``arranged'' coincidence.

Note also recent experimental evidence in favour of
\emph{quark-gluon liquid} in high-energy collisions
\cite{QuarkLiquid} (rather than quark plasma expected in usual
theory) providing an additional, specific support for the proposed
world construction and gravitational protofield structure.

(2) Causal \emph{\textbf{renormalisation of Planckian units}} and
\emph{\textbf{explanation of observed particle spectrum}} (see
also near the end of section \ref{Sec:UniDynEmerge} and
refs.~\cite{Kir:USciCom,Kir:ComDynGrav,Kir:Cosmo}). The important
\emph{practical consequence} here should be an essential change of
accelerator research strategy and purposes (see also item 7
below), as well as serious modification of various cosmology,
particle and quantum physics aspects where Planckian units play an
important role. It includes the predicted absence of notorious
Higgs particles, used in usual theory as a simplified unitary
substitute for the missing property of mass (where a
characteristic \emph{trick} of mechanistic knowledge is performed
in the form of substitution of an artificially invented, abstract
\emph{entity} for a tangibly present, universal \emph{property},
in contradiction with Occam's principle of parsimony and
elementary consistency demand). Since we have \emph{explicitly
obtained} the intrinsic, complex-dynamic \emph{property of mass},
the resource-consuming search for fictitious particles becomes
definitely obsolete, while showing increasingly its practical
uselessness already within conventional approach. The same
actually refers to many other unitary substitutes for the missing
fundamental properties (such as multiple \emph{unobserved}
entities appearing in abstract ``unification'' schemes of
supersymmetry and string theory), which substantiates and
specifies the \emph{inevitable, qualitative change} of the whole
research strategy in high-energy physics.

(3) \emph{\textbf{Causal classicality emergence and macroscopic
quantum effects}}. Our \emph{purely dynamic, intrinsic} mechanism
of emergence of classical, permanently localised type of behaviour
in an elementary \emph{closed} system, in the form of a \emph{new
level} of \emph{unreduced dynamic complexity}
\cite{Kir:USciCom:ArXiv,Kir:75MatWave,Kir:100Quanta,Kir:75Wavefunc,
Kir:QuMach,Kir:USciCom,Kir:Fractal:2,Kir:ComDynGrav,Kir:Cosmo}
(see the end of section \ref{Sec:UniDynEmerge}), is confirmed by,
and practically important for, both transitions from quantum to
classical behaviour and occasional ``revival'' of ``quantum'',
undular behaviour in large enough, normally classical systems
under special conditions.

The causally complete understanding of inevitable transition from
quantum to classical behaviour already on atomic scales is
important for applications like \emph{nanotechnology} and
\emph{many-body quantum devices} that involve complicated and
``entangled'' enough system structure
\cite{Kir:QuMach,Kir:USciCom}, where ``quantum measurement'' and
related classicality emergence \emph{cannot be separated} any more
from ``quantum'' dynamics as such and especially expressed by
contradictory and ``inexplicable'' postulates (as it was still
possible for simplest systems, though at the expense of essential
loss in consistency). Dominating rejection of realistic, causally
complete understanding in those fields produces a huge misleading
effect \cite{Kir:QuMach}.

The ``reverse'' \emph{transition to quantum behaviour} in big
enough, many-body systems is also rich in practically important
applications and needs the unreduced problem solution of quantum
field mechanics, without which the ``clash'' of inconsistent
imitations reveals only ever deepening impasse of unitary
approach. In particular, a popular interpretation of classical
behaviour in terms of ambiguous ``decoherence'' enters in
fundamental contradiction with abundance and diversity of
``quantum revival'' effects that find, on the contrary, their
natural explanation in our intrinsic classicality version
\cite{Kir:USciCom:ArXiv,Kir:75MatWave,Kir:100Quanta,
Kir:75Wavefunc,Kir:QuMach,Kir:USciCom,Kir:Fractal:2,Kir:ComDynGrav,Kir:Cosmo}
(here again, unitary theory tries to reduce intrinsic property to
extrinsic effect from abstract entities).

(4) \emph{\textbf{Realistic quantum machines and causally
substantiated nanotechnology}}. These quickly growing and
extremely versatile applications involve the previous item
aspects, but actually realise, due to arbitrary interaction
structure, unified, many-sided \emph{verification of the whole
micro-physics theory} and ``paradigm''. The \emph{necessity of
change} towards the \emph{unreduced description} of quantum field
mechanics (universal science of complexity) becomes thus
especially evident (see \cite{Kir:QuMach,Kir:Nano} for more
details). We can mention the consistent theory of \emph{quantum
(and classical) chaos}
\cite{Kir:QuMach,Kir:USciCom,Kir:QuChaos,Kir:Channel} and
\emph{quantum measurement} \cite{Kir:USciCom,Kir:QuMeasure} as
examples of essential constituents of the emerging \emph{unified}
description inevitably absent in the unitary theory framework.

(5) \emph{\textbf{Quantum many-body problems with strong
interaction}}. In relation to items~3--4, this group of
applications derives from ``canonical'' studies, but where
many-body systems of interest do not allow any more for usual
application of perturbative approaches and necessitate the
unreduced interaction analysis just provided by the quantum field
mechanics (at the lowest complexity levels). Indeed, it is easy to
see \cite{Kir:QuMach,Kir:USymCom,Kir:SelfOrg,Kir:Fractal:2,
Kir:Nano,Kir:Conscious,Kir:CommNet} that an arbitrary many-body
problem can be expressed by the same system of equations,
eqs.~(\ref{Eq:ExistSystem}), as the protofield interaction problem
considered here (it should be expected physically). Characteristic
applications of interest include complicated solid-state systems
with ``strong'' interaction and ``correlated electrons''
(high-temperature superconductivity, quantum Hall effect, etc.),
quantum atomic condensates and other ``macroscopic quantum
effects'', and causally complete understanding of nuclear dynamics
(including \emph{true} quantum chaos and nuclear stability
estimate \cite{Kir:100Quanta}). It appears, without any surprise,
that such ``difficult'' cases of the ``old'' many-body problem can
be successfully analysed only with the help of its unreduced,
truly non-perturbative general solution for the \emph{arbitrary
interaction} case, which is uniquely provided by the unreduced EP
method (section \ref{Sec:UniDynEmerge}) and contains strong
\emph{qualitative novelties} with respect to any imitation of
unitary, perturbative theory.

(6) \emph{\textbf{Causally complete description of intrinsically
creative cosmological evolution}}. As conventional, unitary theory
does not provide any \emph{creativity} in principle (cannot
explicitly describe autonomous structure formation process), its
application to cases, such as cosmology, with explicit and
``strong'' structure emergence produces the highest number of
mistakes and confusions. Due to ``intrinsically cosmological''
nature of quantum field mechanics (see the beginning of section
\ref{Sec:UniDynEmerge}), it ``automatically'' provides
\emph{explicitly emerging} universe structures, in their
unreduced, realistic version, and thus naturally resolves, or even
does not contain, many ``difficult'', old and new, problems of
unitary cosmology, such as the problems of space-time flatness,
``wavefunction of the universe'' (classical structure emergence),
time flow (entropy), ``quantum cosmology'' (including ``quantum
gravity''), etc. The genuine, complex-dynamical basis of
``relativistic'' behaviour
\cite{Kir:75MatWave,Kir:100Quanta,Kir:75Wavefunc,
Kir:QuMach,Kir:USciCom,Kir:DoubleSol:2,Kir:ComDynGrav,Kir:Cosmo},
outlined in section \ref{Sec:UniDynEmerge}, is naturally involved
with the intrinsically creative, dynamical framework of the new
cosmology, as opposed to the mechanistically ``geometric'' and
abstract imitations of unitary ``models'' that do not accept any
explicit structure formation in principle.

Growing ``new'' problems of dark mass and dark energy, which seem
to defy scandalously conservation of energy and matter itself,
only summarise the \emph{inherent deficiency} (``darkness'') of
unitary, \emph{dynamically single-valued} models used to simulate
the \emph{dynamically multivalued reality}: ``missing'' entities
correspond to the \emph{majority} of essential system realisations
artificially excluded from unitary ``models''. The underlying
``official'' thesis of universe appearance ``from nothing'' is
overthrown by the unreduced interaction description, revealing the
internal \emph{complexity transformation} from dynamic information
to entropy as the unique form of \emph{universal symmetry
(conservation) of complexity} determining any real system
creation, evolution and dynamics
\cite{Kir:QuMach,Kir:USciCom,Kir:USymCom,Kir:Fractal:2,
Kir:Conscious,Kir:CommNet,Kir:SustTrans,Kir:Cosmo}. The
\emph{exact} (never broken) \emph{symmetry} of complexity is the
real basis of essential, causally specified world \emph{progress},
from two coupled homogeneous protofields to the fine-structured,
conscious universe, while nothingness can produce only darkness,
in science and in a real world.

(7) And finally, one should say several words about
\emph{\textbf{direct theory verification}} by ``special''
experiments. The problem with the \emph{quantum world} case is
that one deals here with the \emph{lowest, intrinsically
coarse-grained} (or ``quantum'') levels of complex world dynamics,
which necessarily implies balancing at the border of
\emph{quantitative} resolution and ambiguous ``interpretations''
of technically sophisticated experiments and their faint results
(as opposed to clear and unified \emph{qualitative} manifestations
of unreduced complexity, emphasised by the quantum field
mechanics). While the problems with interpretation of ``fine''
quantum experiments are well known and do not seem to decrease in
usual science framework, we can still try to complete our picture
with discussion of some experimental possibilities of that kind.

Because of fundamental, dynamically obtained quantization of
unreduced interaction process, it is \emph{impossible} to
``catch'' the quantum beat dynamics in any its ``intermediate''
stage or continuous development. However, one can think about some
\emph{indirect}, but special enough, effects of that dynamics
related, in particular, to characteristic, internal resonances at
certain parameter values.

Thus, de Broglie wavelength $\lambda _{\rm{B}}  = {h
\mathord{\left/ {\vphantom {h {mv}}} \right.
\kern-\nulldelimiterspace} {mv}}$ becomes equal to the Compton
wavelength $\lambda _{\rm{C}} = {h \mathord{\left/ {\vphantom {h
{m_0 c}}} \right. \kern-\nulldelimiterspace} {m_0 c}}$ at a total
energy value $E_{\rm{r}}  = \sqrt 2 E_0 $ falling within
accessible acceleration range ($E_{\rm{r}} - E_0 = \left( {\sqrt 2
- 1} \right)E_0 \approx 0,414E_0  = 211 \ {\rm{keV}}$ for the
electron and 389 MeV for proton). Since $\lambda _{\rm{C}}$ is the
size of purely random virtual soliton jumps around the
average-tendency wavelength $\lambda _{\rm{B}}$, their coincidence
for the \emph{unified} quantum beat dynamics can produce an
observable resonant effect at $E = E_{\rm{r}}$ (e.\,g. image
sharpness change in the electron microscope at $E - E_0 \approx
211 \ {\rm{keV}}$).\footnote{This resonance can also be
interpreted as equality between de Broglie frequency, $\nu
_{\rm{B}}$, of regular quantum beat tendency and ``Compton
frequency'' of its totally irregular tendency, $\nu _{\rm{C}}  =
{\mathcal N} = \nu _0 \sqrt {\vphantom{v^2} 1 - {v^2 / c^2}}
\thinspace $, with equal energy partition between the tendencies
(see
eqs.~(\ref{Eq:MotionEnergy}),(\ref{Eq:EnergyPartition})--(\ref{Eq:DeBroglieFrequency})).
Another possibility is the resonant equality of de Broglie
wavelength $\lambda _{\rm{B}}$ and relativistically
\emph{increased} Compton wavelength, ${\mit \Lambda} _{\rm{C}}  =
{c \mathord{\left/ {\vphantom {c {\nu _{\rm{C}} }}} \right.
\kern-\nulldelimiterspace} {\nu _{\rm{C}} }} = \lambda _{\rm{C}} /
\sqrt {\vphantom{v^2} 1 - {v^2 / c^2}} \thinspace $, occurring at
the ``golden mean'' value of particle velocity, ${v
\mathord{\left/ {\vphantom {v c}} \right.
\kern-\nulldelimiterspace} c} = (\sqrt 5  - 1)/ 2 \approx 0,618$
and $E_{\rm{r}}  - E_0  \approx 0,272E_0$.} Similar kind of
resonance can be expected for absorption or emission of photons
(or other exchange particles) with the resonant frequency (or
wavelength) coinciding with those for the average, ``de Broglie''
tendency $\left( \nu _{\rm{B}},\lambda _{\rm{B}} \right)$ or
random, ``Compton'' tendency $\vphantom{v^2} {( \nu _{\rm{C}} =
\nu _0 \sqrt {1 - {v^2 / c^2}} \thinspace , {\mit \Lambda}
_{\rm{C}} = \lambda _{\rm{C}} / \displaystyle \sqrt {1 - {v^2 /
c^2}} \thinspace )}$ for a moving particle
(eqs.~(\ref{Eq:EnergyPartition})--(\ref{Eq:DeBroglieFrequency})).

One more possibility of resonance involving quantum beat dynamics
is particle bound motion or excitation with frequencies attributed
to one, quasi-``photonic'' particle realisation during the quantum
jump phase, $\nu _{\rm{r}}  = {{\nu _0 } \mathord{\left/
{\vphantom {{\nu _0 } {N_\Re  }}} \right.
\kern-\nulldelimiterspace} {N_\Re  }}$, which corresponds to
exchange (e.\,g. photon) energy $h\nu _{\rm{r}}  = {{m_0 c^2 }
\mathord{\left/ {\vphantom {{m_0 c^2 } {N_\Re  }}} \right.
\kern-\nulldelimiterspace} {N_\Re  }} \approx 3726 \ {\rm{eV}}$
for the electron (if $N_\Re   = {{1} \mathord{\left/ {\vphantom
{{1} \alpha }} \right. \kern-\nulldelimiterspace} \alpha } \approx
137$, and one can find the actual $N_\Re$ value).

At a higher complexity sublevel, quantum chaos effects in the
scattering of swift charged particles in crystals should have a
variety of observable manifestations (see ref.~\cite{Kir:Channel}
for details). A very interesting combination of those \emph{two}
quantum complexity sublevels was used for the \emph{first
experimental detection of quantum beat pulsation} by resonance
between ${v \mathord{\left/ {\vphantom {v {\nu _{\rm{C}} }}}
\right. \kern-\nulldelimiterspace} {\nu _{\rm{C}} }} \simeq {\mit
\Lambda} _{\rm{C}}$ for channeled relativistic electrons and
crystal atom spacing \cite{Gouanere}.

All those possibilities can provide \emph{convincing experimental
confirmation} of the quantum field mechanics picture, essentially
exceeding the ambiguity of usual ``interpretational'' experiments
but also promising \emph{novel practical applications}, involving
new ways of particle/energy transformation and structure analysis
at a ``subquantum'' level. A more direct observation of virtual
soliton wandering within its wave is not impossible either, but
can be more problematic in its technical realisation.

In summary, the presented practical implications of the quantum
field mechanics, items (1)--(7), seem to provide enough motivation
for further development of the theory and related experiment. A
real obstacle for further progress comes rather from the
subjective rigidity of traditional, unitary science framework that
does not want even to notice any true novelty, despite the evident
degradation of its own methods and results. Successful completion
of the present project will hopefully contribute to positive shift
towards the unreduced creativity in both spirit and content of
fundamental physics.

}
\centerline{\rule{0pt}{19pt}\rule{72pt}{0.4pt}}

\end{document}